\documentclass[letterpaper,english,conference, twocolumn]{IEEEtran}
\usepackage[T1]{fontenc}
\usepackage[latin9]{inputenc}
\usepackage{babel}
\usepackage{amsthm}
\usepackage{amsmath}
\usepackage{amssymb}
\usepackage{graphicx}
\usepackage{esint}
\usepackage[unicode=true,
 bookmarks=false,
 breaklinks=false,pdfborder={0 0 0},backref=false,colorlinks=false]
 {hyperref}
\hypersetup{pdftitle={Approximately Optimal Trajectory Tracking for an Uncertain Nonlinear System},
 pdfauthor={Rushikesh Kamalapurkar},
 pdfkeywords={Adaptive Dynamic Programming, Approximate Dynamic Programming, Nonlinear Control, Optimal Control},
 pdfstartview= FitBH}
\usepackage{breakurl}

\makeatletter

\providecommand{\tabularnewline}{\\}

\theoremstyle{definition}
\newtheorem{assumption}{Assumption}
  \theoremstyle{plain}
  \newtheorem{thm}{\protect\theoremname}
  \theoremstyle{remark}
  \newtheorem{rem}{\protect\remarkname}

\usepackage{cite}
\IEEEoverridecommandlockouts

\makeatother

\providecommand{\remarkname}{Remark}
\providecommand{\theoremname}{Theorem}

\begin{document}

\title{Approximate optimal cooperative decentralized control for consensus
in a topological network of agents with uncertain nonlinear dynamics%
\thanks{Rushikesh Kamalapurkar, Huyen Dinh, Patrick Walters, and Warren Dixon
are with the Department of Mechanical and Aerospace Engineering, University
of Florida, Gainesville, FL, USA. Email: \{rkamalapurkar, huyentdinh,
walters8, wdixon\}@ufl{}.edu.%
}%
\thanks{This research is supported in part by NSF award numbers 0901491, 1161260,
and 1217908 and ONR grant number N00014-13-1-0151. Any opinions, findings
and conclusions or recommendations expressed in this material are
those of the author(s) and do not necessarily reflect the views of
the sponsoring agency. %
}}

\author{Rushikesh Kamalapurkar, Huyen Dinh, Patrick Walters, and Warren Dixon}
\maketitle
\begin{abstract}
Efforts in this paper seek to combine graph theory with adaptive dynamic
programming (ADP) as a reinforcement learning (RL) framework to determine
forward-in-time, real-time, approximate optimal controllers for distributed
multi-agent systems with uncertain nonlinear dynamics. A decentralized
continuous time-varying control strategy is proposed, using only local
communication feedback from two-hop neighbors on a communication topology
that has a spanning tree. An actor-critic-identifier architecture
is proposed that employs a nonlinear state derivative estimator to
estimate the unknown dynamics online and uses the estimate thus obtained
for value function approximation. Simulation results demonstrate the
applicability of the proposed technique to cooperatively control a
group of five agents.
\end{abstract}

\section{Introduction}

Combined efforts from multiple autonomous agents can yield tactical
advantages including: improved munitions effects; distributed sensing,
detection, and threat response; and distributed communication pipelines.
While coordinating behaviors among autonomous agents is a challenging
problem that has received mainstream focus, unique challenges arise
when seeking autonomous collaborative behaviors in low bandwidth communication
environments. For example, most collaborative control literature focuses
on centralized approaches that require all nodes to continuously communicate
with a central agent, yielding a heavy communication demand that is
subject to failure due to delays, and missing information. Furthermore,
the central agent requires to carry enough computational resources
on-board to process the data and to generate command signals. These
challenges motivate the need for a decentralized approach where the
nodes only need to communicate with their neighbors for guidance,
navigation and control tasks.

Reinforcement learning (RL) allows an agent to learn the optimal policy
by interacting with its environment, and hence, is useful for control
synthesis in complex dynamical systems such as a network of agents.
Decentralized algorithms have been developed for cooperative control
of networks of agents with finite state and action spaces in \cite{Lauer.Riedmiller2000,Busoniu.Babuska.ea2008,Weib1995,Cao.Chen.ea2011}.
See \cite{Busoniu.Babuska.ea2008} for a survey. The extension of
these techniques to networks of agents with infinite state and action
spaces and nonlinear dynamics is challenging due to difficulties in
value function approximation, and has remained an open problem.

As the desired action by an individual agent depends on the actions
and the resulting trajectories of its neighbors, the error system
for each agent becomes a complex nonautonomous dynamical system. Nonautonomous
systems, in general, have non-stationary value functions. As non-stationary
functions are difficult to approximate using parametrized function
approximation schemes such as neural networks (NNs), designing optimal
policies for nonautonomous systems is not trivial. To get around this
challenge, differential game theory is often employed in multi-agent
optimal control, where a solution to the coupled Hamilton-Jacobi-Bellman
(HJB) equation (c.f. \cite{Vamvoudakis2011}) is sought. As the coupled
HJB equations are difficult to solve, some form of generalized policy
iteration or value iteration \cite{Sutton1998} is often employed
to get an approximate solution. It is shown in results such as \cite{Vamvoudakis.Lewis.ea2012,Vamvoudakis2010a,Vamvoudakis2011,Vrabie2010,Johnson2010,Johnson2011a}
that approximate dynamic programming (ADP) can be used to generate
approximate optimal policies online for multi-agent systems. As the
HJB equations to be solved are coupled, all of these results have
a centralized control architecture.

Decentralized control techniques focus on finding control policies
based on local data for individual agents that collectively achieve
the desired goal, which, for the problem considered in this effort,
is consensus to the origin. Various methods have been developed to
solve the consensus problem for linear systems with exact model knowledge.
An optimal control approach is used in \cite{Wang.Xin2010} to achieve
consensus while avoiding obstacles. In \cite{Wang.Xin2012}, an optimal
controller is developed for agents with known dynamics to cooperatively
track a desired trajectory. In \cite{Semsar-Kazerooni.Khorasani2008},
an optimal consensus algorithm is developed for a cooperative team
of agents with linear dynamics using only partial information. A value
function approximation based approach is presented in \cite{Vamvoudakis.Lewis.ea2012a}
for cooperative synchronization in a strongly connected network of
agents with known linear dynamics. It is also shown in \cite{Vamvoudakis.Lewis.ea2012a}
that the obtained policies are in a cooperative Nash equilibrium. 

For nonlinear systems, a model predictive control approach is presented
in \cite{Shim.Kim.ea2003}, however, no stability or convergence analysis
is presented. A stable distributed model predictive controller is
presented in \cite{Magni.Scattolini2006} for nonlinear discrete-time
systems with known nominal dynamics. Asymptotic stability is proved
without any interaction between the nodes, however, a nonlinear optimal
control problem need to be solved at every iteration to implement
the controller. Decentralized optimal control synthesis for consensus
in a topological network of agents with continuous-time uncertain
nonlinear dynamics has remained an open problem.

In this result, an ADP-based approach is developed to solve the consensus
problem for a network topology that has a spanning tree. The agents
are assumed to have nonlinear control-affine dynamics with unknown
drift vectors and known control effectiveness matrices. An identifier
is used in conjunction with the controller enabling the algorithm
to find approximate optimal decentralized policies online without
the knowledge of drift dynamics. This effort thus realizes the actor-critic-identifier
(ACI) architecture (c.f. \cite{Werbos1992,Bhasin.Kamalapurkar.ea2013a})
for networks of agents. Simulations are presented to demonstrate the
applicability of the proposed technique to cooperatively control a
group of five agents.

\section{Graph Theory Preliminaries}

Let $\mathcal{N}\triangleq\left\{ \beta_{1},\beta_{2},\cdots,\beta_{N}\right\} $
denote a set of $N$ agents moving in the state space $S\subset\mathbb{R}^{n}$.
The objective is for the agents to reach a consensus state. Without
loss of generality, let the consensus state be the origin of the state
space, i.e. $S\owns x_{0}=0$. To aid the subsequent design, the agent
$\beta_{0}$ (henceforth referred to as the leader) is assumed to
be stationary at the origin. The agents are assumed to be on a network
with a fixed communication topology modeled as a static directed graph
(i.e. digraph). 

Each agent forms a node in the digraph. If agent $\beta_{j}$ can
communicate with agent $\beta_{i}$ then there exists a directed edge
from the $j^{th}$ to the $i{}^{th}$ node of the digraph, denoted
by the ordered pair $\left(\beta_{j},\beta_{i}\right)\in\mathcal{N}\times\mathcal{N}$.
Let $E\subset\mathcal{N}\times\mathcal{N}$ denote the set of all
edges. Let there be a positive weight $a_{ij}\in\mathbb{R}$ associated
with each edge $\left(\beta_{j},\beta_{i}\right)$. Note that $a_{ij}\neq0$
if and only if $\left(\beta_{j},\beta_{i}\right)\in E.$ The digraph
is assumed to have no repeated edges i.e. $\left(\beta_{i},\beta_{i}\right)\notin E,\forall i$,
which implies $a_{ii}=0,\forall i$. Note that $a_{i0}$ denotes the
edge weight (also referred to as the pinning gain) for the edge between
the leader and an agent $\beta_{i}$. Similar to the other edge weights,
$a_{i0}\neq0$ if and only if there exists a directed edge from the
leader to the agent $i$. The neighborhood set of agent $\beta_{i}$
is denoted by $\mathcal{N}_{i}$ defined as $\mathcal{N}_{i}\triangleq\left\{ j\mid\left(\beta_{j},\beta_{i}\right)\in E\right\} $.
To streamline the analysis, the graph connectivity matrix $\mathcal{A}\in\mathbb{R}^{N\times N}$
is defined as $\mathcal{A}\triangleq\left[a_{ij}\mid i,j=1,\cdots,N\right]$,
the pinning gain matrix $\mathcal{A}_{0}\in\mathbb{R}^{N\times N}$
is a diagonal matrix defined as $\mathcal{A}_{0}\triangleq diag\left(a_{i0}\right)\mid i=1,\cdots,N$,
the matrix $\mathcal{D}\in\mathbb{R}^{N\times N}$ is defined as $\mathcal{D}\triangleq diag\left(d_{i}\right),$
where $d_{i}\triangleq\sum_{j\in\mathcal{N}_{i}}a_{ij}$, and the
graph Laplacian matrix $\mathcal{L}\in\mathbb{R}^{N\times N}$ is
defined as $\mathcal{L}\triangleq\mathcal{D}-\mathcal{A}$. The graph
is said to have a spanning tree if given any node $\beta_{i}$, there
exists a directed path from the leader $\beta_{0}$ to $\beta_{i}$.
For notational brevity, a linear operator $\Upsilon_{i}\left(\left(\cdot\right)\right)$
is defined as 
\begin{align}
\Upsilon_{i}\left(\left(\cdot\right)\right) & \triangleq\left(\sum_{j\in\mathcal{N}_{i}}a_{ij}\left(\left(\cdot\right)_{i}-\left(\cdot\right)_{j}\right)+a_{i0}\left(\left(\cdot\right)_{i}\right)\right).\label{eq:uPSILON}
\end{align}

\section{Problem Definition}

Let the dynamics of each agent be described as 
\[
\dot{x}_{i}=f_{i}\left(x_{i}\right)+g_{i}\left(x_{i}\right)u_{i},\forall i=1,2,\cdots,N
\]
where $x_{i}\left(\cdot\right)\in S\subset\mathbb{R}^{n}$ is the
state, $f_{i}:S\to\mathbb{R}^{n}$ and $g_{i}:S\to\mathbb{R}^{n\times m}$
are locally Lipschitz functions, and $u_{i}\left(\cdot\right)\subset\mathbb{R}^{m}$
is the control policy. To achieve consensus to the leader, define
the local neighborhood tracking error $e_{i}\left(\cdot\right)\in S\subset\mathbb{R}^{n}$
for each agent as \cite{Khoo.Xie2009}
\begin{equation}
e_{i}\triangleq\Upsilon_{i}\left(x\right)=\sum_{j\in\mathcal{N}_{i}}a_{ij}\left(x_{i}-x_{j}\right)+a_{i0}\left(x_{i}\right).\label{eq:e}
\end{equation}

Denote the cardinality of the set $\mathcal{N}_{i}$ by $\left|\mathcal{N}_{i}\right|$.
Let $\mathcal{E}_{i}\left(\cdot\right)\in S^{\left|\mathcal{N}_{i}\right|+1}\subseteq\mathbb{R}^{\left|\mathcal{N}_{i}\right|+1}$
be a stacked vector of local neighborhood tracking errors corresponding
to the agent $\beta_{i}$ and its neighbors, i.e., $\mathcal{E}_{i}\triangleq\left\{ e_{j}\mid j\in\mathcal{N}_{i}\right\} \cup\left\{ e_{i}\right\} $.
To achieve consensus in an optimal cooperative way, it is desired
to minimize, for each agent, the cost $J_{i}\triangleq\frac{1}{2}\intop_{0}^{\infty}r_{i}\left(\mathcal{E}_{i},u_{i}\right)dt,$
where 
\begin{equation}
r_{i}\left(\mathcal{E}_{i},u_{i}\right)\triangleq e_{i}^{T}Q_{ii}e_{i}+u_{i}^{T}R_{i}u_{i}+\sum_{j\in\mathcal{N}_{i}}a_{ij}e_{j}^{T}Q_{ij}e_{j}.\label{eq:ri}
\end{equation}
In (\ref{eq:ri}), $R_{i}\in\mathbb{R}^{m\times m}$ and $Q_{ii},Q_{ij}\in\mathbb{R}^{n\times n}$
are symmetric positive definite matrices of constants. Let $\mathcal{E}\triangleq\begin{bmatrix}e_{1}^{T} & e_{2}^{T} & \cdots & e_{N}^{T}\end{bmatrix}^{T}\in S^{nN}\subset\mathbb{R}^{nN}$
and $\mathcal{X}\triangleq\begin{bmatrix}x_{1}^{T} & x_{2}^{T} & \cdots & x_{N}^{T}\end{bmatrix}^{T}\in S^{nN}\subset\mathbb{R}^{nN}$.
Using the definition of $e_{i}$ from (\ref{eq:e}) we get 
\[
\mathcal{E}=\left[\begin{array}{c}
\Upsilon_{1}\left(x\right)\\
\Upsilon_{2}\left(x\right)\\
\vdots\\
\Upsilon_{N}\left(x\right)
\end{array}\right]=\left(\left(\mathcal{L}+\mathcal{A}_{0}\right)\otimes I_{n}\right)\mathcal{X},
\]
where $\otimes$ denotes the Kronecker product and $I_{n}\in\mathbb{R}^{n\times n}$
is the identity matrix.

\section{Control Development}

\subsection{State derivative estimation\label{sub:State-derivative-estimation}}

Based on the development in \cite{Bhasin.Kamalapurkar.ea2013a}, each
agent's dynamics can be approximated using a dynamic neural network
(DNN) with $M_{fi}$ hidden layer neurons as
\[
\dot{x}_{i}=W_{fi}^{T}\sigma(V_{fi}^{T}x_{i})+\varepsilon_{fi}(x_{i})+g_{i}(x_{i})u_{i},
\]
where $W_{fi}\in\mathbb{R}^{M_{fi}+1\times n}$, $V_{fi}\in\mathbb{R}^{n\times M_{fi}}$
are unknown ideal DNN weights, $\sigma_{fi}\triangleq\sigma(V_{fi}^{T}x_{i})\in\mathbb{R}^{M_{fi}+1}$
is a bounded DNN activation function, and $\varepsilon_{fi}:\mathbb{R}^{n}\to\mathbb{R}^{n}$
is the function reconstruction error. In the following, the drift
dynamics $f_{i}$ are unknown and the control effectiveness functions
$g_{i}$ are assumed to be known. Each agent estimates the derivative
of its own state using the following state-derivative estimator
\begin{align}
\dot{\hat{x}}_{i} & =\hat{f}_{i}+g_{i}(x_{i})u_{i}+\mu_{i},\quad\hat{f}_{i}\triangleq\hat{W}_{fi}^{T}\hat{\sigma}_{fi},\nonumber \\
\dot{\hat{W}}_{fi} & =proj(\Gamma_{wfi}\hat{\sigma}_{fi}'\hat{V}_{fi}^{T}\dot{\hat{x}}_{i}\tilde{x}_{i}^{T}),\nonumber \\
\dot{\hat{V}}_{fi} & =proj(\Gamma_{vfi}\dot{\hat{x}}_{i}\tilde{x}_{i}^{T}\hat{W}_{fi}^{T}\hat{\sigma}_{fi}'),\nonumber \\
\mu_{i} & \triangleq k_{fi}\tilde{x}_{i}(t)-k_{fi}\tilde{x}_{i}(0)+v_{i},\nonumber \\
\dot{v}_{i} & =(k_{fi}\alpha_{fi}+\gamma_{fi})\tilde{x}_{i}+\beta_{1fi}sgn(\tilde{x}_{i}),\quad v_{i}\left(0\right)=0,\label{eq:nu}
\end{align}
where $\hat{W}_{fi}\left(\cdot\right)\in\mathbb{R}^{M_{fi}+1\times n}$
and $\hat{V}_{fi}\left(\cdot\right)\in\mathbb{R}^{M_{fi}+1\times n}$
are the estimates for the ideal DNN weights $W_{fi}$ and $V_{fi}$,
$\hat{x}_{i}\left(\cdot\right)\in\mathbb{R}^{n}$ is the state estimate,
$\hat{\sigma}_{fi}\triangleq\sigma(\hat{V}_{fi}^{T}\hat{x}_{i})\in\mathbb{R}^{M_{fi}+1}$,
$\tilde{x}_{i}\triangleq x_{i}-\hat{x}_{i}\in\mathbb{R}^{n}$ is the
state estimation error, $k_{fi},\:\alpha_{fi},\:\gamma_{fi},\:\beta_{1fi}\in\mathbb{R}$
are positive constant control gains, $proj\left\{ \cdot\right\} $
is a smooth projection operator \cite{Dixon2003}, and $v_{i}\left(\cdot\right)\in\mathbb{R}^{n}$
is a generalized Filippov solution to (\ref{eq:nu}). For notational
brevity define 
\begin{align*}
\hat{F}_{i}\left(x_{i},\hat{x}_{i},u_{i},t\right) & \triangleq\hat{f}_{i}\left(\hat{x}_{i}\right)+g_{i}\left(x_{i}\right)u_{i}+\mu_{i}\left(t\right),\\
F_{i}\left(x_{i},u_{i}\right) & \triangleq f_{i}\left(x_{i}\right)+g_{i}\left(x_{i}\right)u_{i},\quad\tilde{F}_{i}\triangleq\hat{F}_{i}-F_{i}.
\end{align*}
It is shown in \cite[Theorem 1]{Bhasin.Kamalapurkar.ea2013a} that
provided the gains $k_{fi}$ and $\gamma_{fi}$ are sufficiently large
and $x_{i}$ and $u_{i}$ are bounded, the estimation error $\tilde{x}_{i}$
and its derivative are bounded. Furthermore, $\lim_{t\to\infty}\left\Vert \tilde{x}_{i}\left(t\right)\right\Vert =0,$
$\lim_{t\to\infty}\left\Vert \dot{\tilde{x}}_{i}\left(t\right)\right\Vert =0,$
and $\tilde{F}_{i}\in\mathcal{L}_{\infty}.$

\subsection{Value function approximation}

The value function $V_{i}:S^{\left|\mathcal{N}_{i}\right|+1}\to\mathbb{R}^{+}$
is the cost-to-go for each agent given by 
\begin{equation}
V_{i}\left(\mathcal{E}_{i}^{o}\right)=\frac{1}{2}\intop_{t_{0}}^{\infty}r_{i}\left(\mathcal{E}_{i}\left(\tau\right),u_{i}\left(\mathcal{E}_{i}\left(\tau\right)\right)\right)d\tau,\label{eq:Vi}
\end{equation}
where $\mathcal{E}_{i}\left(\tau\right)$ denote the neighborhood
tracking error trajectories associated with agent $\beta_{i}$ and
its neighbors, with the initial conditions $\mathcal{E}_{i}\left(t_{0}\right)=\mathcal{E}_{i}^{o}.$
The time derivative of $V_{i}$ is then given by 
\[
\dot{V}_{i}=\sum_{j\in i\cup\mathcal{N}_{i}}\frac{\partial V_{i}}{\partial e_{j}}\Upsilon_{j}\left(F\right).
\]
The Hamiltonian for the optimal control problem is the differential
equivalent of $\left(\ref{eq:Vi}\right)$ given by
\begin{align*}
H_{i} & \triangleq r_{i}+\sum_{j\in i\cup\mathcal{N}_{i}}\frac{\partial V_{i}}{\partial e_{j}}\Upsilon_{j}\left(F\right).
\end{align*}
The optimal value function $V_{i}^{*}:S^{\left|\mathcal{N}_{i}\right|+1}\to\mathbb{R}^{+}$
is defined as 
\begin{equation}
V_{i}^{*}\left(\mathcal{E}_{i}^{o}\right)\triangleq\underset{u_{i}\in U_{i}}{\underset{u_{i}:S^{\left|\mathcal{N}_{i}\right|+1}\to\mathbb{R}^{m}}{min}}\frac{1}{2}\intop_{t}^{\infty}r_{i}\left(\mathcal{E}_{i}\left(\tau\right),u_{i}\left(\mathcal{E}_{i}\left(\tau\right)\right)\right)d\tau,\label{eq:Vi*}
\end{equation}
where $U_{i}$ denotes the set of all admissible policies for the
agent $\beta_{i}$ \cite{Beard1997}. Assuming that the minimizer
in (\ref{eq:Vi*}) exists, $V_{i}^{*}$ is the solution to the HJB
equation
\begin{equation}
H_{i}^{*}=r_{i}^{*}\left(\mathcal{E}_{i},u_{i}^{*}\right)+\sum_{j\in i\cup\mathcal{N}_{i}}\frac{\partial V_{i}^{*}}{\partial e_{j}}\Upsilon_{j}\left(F^{*}\right)=0,\label{eq:Hi*}
\end{equation}
where $F_{i}^{*}\left(x_{i},u_{i}^{*}\right)\triangleq f_{i}\left(x_{i}\right)+g_{i}(x_{i})u_{i}^{*}$,
and the minimizer in (\ref{eq:Vi*}) is the optimal policy $u_{i}^{*}:S^{\left|\mathcal{N}_{i}\right|+1}\to\mathbb{R}^{m}$,
which can be obtained by solving the equation $\frac{\partial H_{i}^{*}\left(\mathcal{E}_{i},u_{i}^{*}\right)}{\partial u_{i}^{*}}=0$.
Using the definition of $H_{i}^{*}$ in $\left(\ref{eq:Hi*}\right)$,
the optimal policy can be written in a closed form as
\begin{equation}
u_{i}^{*}=-\frac{1}{2}R_{i}^{-1}g_{i}^{T}\left(\left(a_{i0}+d_{i}\right)\left(V_{ie_{i}}^{*}\right)^{T}-\sum_{j\in\mathcal{N}_{i}}a_{ji}\left(V_{ie_{j}}^{*}\right)^{T}\right),\label{eq:u*}
\end{equation}
where $V_{ie_{i}}^{*}\triangleq\frac{\partial V_{i}^{*}}{\partial e_{i}}$,
and $V_{ie_{j}}^{*}\triangleq\frac{\partial V_{i}^{*}}{\partial e_{j}}$,
assuming that the optimal value function $V_{i}^{*}$ satisfies $V_{i}^{*}\in C^{1}$
and $V_{i}^{*}\left(0\right)=0.$ Note that the controller for node
$i$ only requires the tracking error and edge weight information
from itself and its neighbors. The following assumptions are made
to facilitate the use of NNs to approximate the optimal policy and
the optimal value function.
\begin{assumption}
\label{Acompact} The set $S$ is compact. Based on the subsequent
stability analysis, this assumption holds as long as the initial condition
$\mathcal{X}\left(0\right)$ is bounded. See Remark \ref{rem:chicompact}
in the subsequent stability analysis. 
\end{assumption}

\begin{assumption}
\label{ANN}Each optimal value function $V_{i}^{*}$ can be represented
using a NN with $M_{i}$ neurons as
\begin{equation}
V_{i}^{*}\left(\mathcal{E}_{i}\right)=W_{i}^{T}\sigma_{i}\left(\mathcal{E}_{i}\right)+\epsilon_{i}\left(\mathcal{E}_{i}\right),\label{eq:V*NN}
\end{equation}
where $W_{i}\in\mathbb{R}^{M_{i}}$ is the ideal weight matrix bounded
above by a known positive constant $\bar{W}_{i}\in\mathbb{R}$ in
the sense that $\left\Vert W_{i}\right\Vert _{2}\leq\bar{W}_{i}$,
$\sigma_{i}:S^{\left|\mathcal{N}_{i}\right|+1}\rightarrow\mathbb{R}^{M_{i}}$
is a bounded continuously differentiable nonlinear activation function,
and $\epsilon_{i}:S^{\left|\mathcal{N}_{i}\right|+1}\rightarrow\mathbb{R}$
is the function reconstruction error such that $\sup_{\mathcal{E}_{i}}\left|\epsilon_{i}\left(\mathcal{E}_{i}\right)\right|\leq\bar{\epsilon}_{i}$
and\textbf{ }$\sup_{\mathcal{E}_{i}}\left\Vert \epsilon_{i}'\left(\mathcal{E}_{i}\right)\right\Vert \leq\bar{\epsilon'}_{i}$,
where $\epsilon_{i}'=\frac{\partial\epsilon_{i}}{\partial\mathcal{E}_{i}}$
and $\bar{\epsilon}_{i},\:\bar{\epsilon'}_{i}\in\mathbb{R}$ are positive
constants \cite{Hornik1990,Lewis2002}. 
\end{assumption}
From $\left(\ref{eq:u*}\right)$ and $\left(\ref{eq:V*NN}\right)$
the optimal policy can be represented as
\begin{equation}
u_{i}^{*}=-\frac{1}{2}R_{i}^{-1}g_{i}^{T}\left(L_{\sigma i}W_{i}+L_{\epsilon i}\right),\label{eq:mu*NN}
\end{equation}
where $L_{\sigma i}\triangleq\left(\left(a_{i0}+d_{i}\right)\left(\frac{\partial\sigma_{i}}{\partial e_{i}}\right)^{T}-\sum_{j\in\mathcal{N}_{i}}a_{ji}\left(\frac{\partial\sigma_{i}}{\partial e_{j}}\right)^{T}\right),$
and $L_{\epsilon i}\triangleq\left(\left(a_{i0}+d_{i}\right)\left(\frac{\partial\epsilon_{i}}{\partial e_{i}}\right)^{T}-\sum_{j\in\mathcal{N}_{i}}a_{ji}\left(\frac{\partial\epsilon_{i}}{\partial e_{j}}\right)^{T}\right).$

Based on $\left(\ref{eq:V*NN}\right)$ and $\left(\ref{eq:mu*NN}\right)$,
the NN approximations to the optimal value function and the optimal
policy are given by
\begin{gather}
\hat{V}_{i}=\hat{W}_{ci}^{T}\sigma_{i},\quad u_{i}=-\frac{1}{2}R_{i}^{-1}g_{i}^{T}L_{\sigma i}\hat{W}_{ai},\label{eq:muhat}
\end{gather}
where $\hat{W}_{ci}\left(\cdot\right)\in\mathbb{R}^{M_{i}}$ and $\hat{W}_{ai}\left(\cdot\right)\in\mathbb{R}^{M_{i}}$
are estimates of the ideal neural network weights $W_{i}$. Using
$\left(\ref{eq:V*NN}\right)$-$\left(\ref{eq:muhat}\right)$, the
approximate Hamiltonian $\hat{H}_{i}\left(\cdot\right)$ and the optimal
Hamiltonian $H_{i}^{*}\left(\cdot\right)$ can be obtained as
\begin{align}
\hat{H}_{i} & =e_{i}^{T}Q_{ii}e_{i}+u_{i}^{T}R_{i}u_{i}+\sum_{j=1}^{N}a_{ij}e_{j}^{T}Q_{ij}e_{j}+\hat{W}_{ci}^{T}\omega_{i},\nonumber \\
H_{i}^{*} & =e_{i}^{T}Q_{ii}e_{i}+u_{i}^{*T}R_{i}u_{i}^{*}+\sum_{j=1}^{N}a_{ij}e_{j}^{T}Q_{ij}e_{j}\nonumber \\
 & +W_{i}^{T}\omega_{i}^{*}+\epsilon_{iF^{*}}^{\prime},\label{eq:HiHat}
\end{align}
where $\epsilon_{iF^{*}}^{\prime}\triangleq\sum_{j\in i\cup\mathcal{N}_{i}}\left(\frac{\partial\epsilon_{i}}{\partial e_{j}}\right)\Upsilon_{j}\left(F^{*}\right)$
and 
\begin{gather}
\omega_{i}\triangleq\sum_{j\in i\cup\mathcal{N}_{i}}\left(\frac{\partial\sigma_{i}}{\partial e_{j}}\right)\Upsilon_{j}\left(\hat{F}\right),\nonumber \\
\omega_{i}^{*}\triangleq\sum_{j\in i\cup\mathcal{N}_{i}}\left(\frac{\partial\sigma_{i}}{\partial e_{j}}\right)\Upsilon_{j}\left(F^{*}\right).\label{eq:omega}
\end{gather}
Using $\left(\ref{eq:Hi*}\right),$ the error between the approximate
and the optimal Hamiltonian, called the Bellman error (BE) $\delta_{i}\left(\cdot\right)\in\mathbb{R}$,
is given in a measurable form by
\begin{gather}
\delta_{i}\triangleq\hat{H}_{i}-H_{i}^{*}=\hat{H}_{i}.\label{eq:delta}
\end{gather}

Note that equations (\ref{eq:HiHat})-(\ref{eq:delta}) imply that
to compute the BE, the $i{}^{th}$ agent requires the knowledge of
$\Upsilon_{i}\left(\hat{F}\right)$ and $\Upsilon_{j}\left(\hat{F}\right)$
for all $j\in\mathcal{N}_{i}$. As each agent can compute its own
$\Upsilon_{i}\left(\hat{F}\right)$ based on local information, the
computation of $\delta_{i}$ for each agent can be achieved via two-hop
local communication.

The primary contribution of this result is that the developed value
function approximation scheme, together with the state derivative
estimator, enables the computation of the BE $\delta_{i}$ with only
local information, and without the knowledge of drift dynamics. Furthermore,
unlike the previous results such as \cite{Vamvoudakis.Lewis.ea2012a},
the effect of the local tracking errors of the neighbors of an agent
is explicitly considered in the HJB equation for that agent, resulting
in the novel control law in (\ref{eq:muhat}). In the following, the
update laws for the value function and the policy weight estimates
based on the BE are presented. The update laws and the subsequent
development leading up to the stability analysis in Section \ref{sec:Stability-Analysis}
are similar to our previous result in \cite{Bhasin.Kamalapurkar.ea2013a}
with minor changes, and are presented here for completeness. 

Note that the BE in (\ref{eq:delta}) is linear in the value function
weight estimates $\hat{W}_{ci}$ and nonlinear in the policy weight
estimates $\hat{W}_{ai}$. The use of two different sets of weights
to approximate the same ideal weights $W_{i}$ is motivated by the
heuristic observation that adaptive update laws based on least squares
minimization perform better than those based on gradient descent.
As the application of least squares technique requires linearity of
the error with respect to the parameters being estimated, the use
of two different sets of weights facilitates the development of a
least squares minimization-based update law for the value function
weights. The value function weights are updated to minimize $\intop_{0}^{t}\delta_{i}^{2}\left(\tau\right)d\tau$
using a least squares update law with a forgetting factor as \cite{Johnstone1982,Ioannou1996}
\begin{align}
\dot{\hat{W}}_{ci} & =-\phi_{ci}\gamma_{i}\frac{\omega_{i}}{1+\nu_{i}\omega_{i}^{T}\gamma_{i}\omega_{i}}\delta_{i},\label{eq:WcHatdot}\\
\dot{\gamma}_{i} & =-\phi_{ci}\left(-\lambda_{i}\gamma_{i}+\gamma_{i}\frac{\omega_{i}\omega_{i}^{T}}{1+\nu_{i}\omega_{i}^{T}\gamma_{i}\omega_{i}}\gamma_{i}\right),\label{eq:Gammadot}
\end{align}
where $\nu_{i},\phi_{ci}\in\mathbb{R}$ are positive adaptation gains,
$\lambda_{i}\in\left(0,1\right)$ is the forgetting factor for the
estimation gain matrix $\gamma_{i}\left(\cdot\right)\in\mathbb{R}^{M_{i}\times M_{i}}$.
The policy weights are updated to follow the value function weight
estimates as
\begin{align}
\overset{\cdot}{\hat{W}}_{ai} & =proj\left\{ -\phi_{ai2}\left(\hat{W}_{ai}-\hat{W}_{ci}\right)\right\} ,\label{eq:WaHatdot}
\end{align}
where $\phi_{ai2}\in\mathbb{R}$ is a positive adaptation gain, and
$proj\left\{ \cdot\right\} $ is a smooth projection operator \cite{Dixon2003}.
The use of forgetting factor ensures that 
\begin{equation}
\underline{\varphi_{i}}I_{M_{i}}\leq\gamma_{i}\left(t\right)\leq\overline{\varphi_{i}}I_{M_{i}},\quad\forall t\in[t_{0},\infty),\label{eq:GammaBound}
\end{equation}
where $\overline{\varphi_{i}},\underline{\varphi_{i}}\in\mathbb{R}$
are constants such that $0<\underline{\varphi_{i}}<\overline{\varphi_{i}}$
\cite{Johnstone1982,Ioannou1996}. Using $\left(\ref{eq:HiHat}\right)$-$\left(\ref{eq:WcHatdot}\right)$,
an unmeasurable form of the BE can be written as
\begin{align}
\delta_{i} & =-\tilde{W}_{ci}^{T}\omega_{i}+\frac{1}{4}\hat{W}_{ai}^{T}G_{\sigma i}\hat{W}_{ai}-\frac{1}{4}W_{i}^{T}G_{\sigma i}W_{i}-\frac{1}{2}W_{i}^{T}G_{\sigma\epsilon i}\nonumber \\
 & +\frac{1}{2}W_{i}^{T}\left(\sum_{j\in i\cup\mathcal{N}_{i}}\sigma_{ie_{j}}\Upsilon_{j}\left(GL_{\sigma}\tilde{W}_{a}+GL_{\epsilon}\right)\right)\nonumber \\
 & +W_{i}^{T}\left(\sum_{j\in i\cup\mathcal{N}_{i}}\sigma_{ie_{j}}\Upsilon_{j}\left(\tilde{F}\right)\right)-\frac{1}{4}G_{\epsilon i}-\epsilon_{iF^{*}}^{\prime},\label{eq:deltaUnm}
\end{align}
The weight estimation errors for the value function and the policy
are defined as $\tilde{W}_{ci}\left(t\right)\triangleq W_{i}-\hat{W}_{ci}\left(t\right)$
and $\tilde{W}_{ai}\left(t\right)\triangleq W_{i}-\hat{W}_{ai}\left(t\right)$,
respectively. Using (\ref{eq:delta}), the weight estimation error
dynamics for the value function can be rewritten as
\begin{align}
\dot{\tilde{W}}_{ci} & =-\phi_{ci}\gamma_{i}\psi_{i}\psi_{i}^{T}\tilde{W}_{ci}+\frac{\phi_{ci}\gamma_{i}\omega_{i}}{1+\nu_{i}\omega_{i}^{T}\gamma_{i}\omega_{i}}\Biggl(\nonumber \\
 & W_{i}^{T}\left(\sum_{j\in i\cup\mathcal{N}_{i}}\sigma_{ie_{j}}\Upsilon_{j}\left(\tilde{F}\right)\right)-\frac{1}{4}G_{\epsilon i}-\epsilon_{iF^{*}}^{\prime}\nonumber \\
 & +\frac{1}{2}W_{i}^{T}\left(\sum_{j\in i\cup\mathcal{N}_{i}}\sigma_{ie_{j}}\Upsilon_{j}\left(GL_{\sigma}\tilde{W}_{a}+GL_{\epsilon}\right)\right)\nonumber \\
 & +\frac{1}{4}\tilde{W}_{ai}^{T}G_{\sigma i}\tilde{W}_{ai}-\frac{1}{2}\tilde{W}_{i}^{T}G_{\sigma i}W_{i}-\frac{1}{2}W_{i}^{T}G_{\sigma\epsilon i}\Biggr),\label{eq:WcDyn}
\end{align}
where $\sigma_{ie_{j}}\triangleq\frac{\partial\sigma_{i}}{\partial e_{j}}$,
$G_{i}\triangleq g_{i}R_{i}^{-1}g_{i}^{T}$, $G_{\sigma i}\triangleq L_{\sigma i}g_{i}R_{i}^{-1}g_{i}^{T}L_{\sigma i}$,
$G_{\epsilon i}\triangleq L_{\epsilon i}g_{i}R_{i}^{-1}g_{i}^{T}L_{\epsilon i}$,
$G_{\sigma\epsilon i}\triangleq L_{\sigma i}g_{i}R_{i}^{-1}g_{i}^{T}L_{\epsilon i}$
and$\psi_{i}\left(\cdot\right)\triangleq\frac{\omega_{i}}{\sqrt{1+\nu_{i}\omega_{i}^{T}\gamma_{i}\omega_{i}}}\in\mathbb{R}^{M_{i}}$
is the regressor vector. Based on $\left(\ref{eq:GammaBound}\right),$
the regressor vector can be bounded as
\begin{equation}
\left\Vert \psi_{i}\left(t\right)\right\Vert \leq\frac{1}{\sqrt{\nu_{i}\underline{\varphi_{i}}}},\quad\forall t\in[t_{0},\infty).\label{eq:psibound}
\end{equation}
The dynamics in $\left(\ref{eq:WcDyn}\right)$ can be regarded as
a perturbed form of the nominal system
\begin{equation}
\dot{\tilde{W}}_{ci}=-\phi_{ci}\gamma_{i}\psi_{i}\psi_{i}^{T}\tilde{W}_{ci}.\label{eq:WcNom}
\end{equation}
Using Corollary 4.3.2 in \cite{Ioannou1996} and Assumption \ref{Acompact},
$\left(\ref{eq:WcNom}\right)$ is globally exponentially stable if
the regressor vector $\psi_{i}:[0,\infty)\to\mathbb{R}^{M_{i}}$ is
persistently exciting. Given $\left(\ref{eq:GammaBound}\right)$,
$\left(\ref{eq:psibound}\right)$, and $\left(\ref{eq:WcNom}\right)$,
Theorem 4.14 in \cite{Khalil2002} can be used to show that there
exists a function $V_{ci}:\mathbb{R}^{M_{i}}\times\left[0,\infty\right)\rightarrow\mathbb{R}$
and positive constants $\underline{v_{ci}}$, $\overline{v_{ci}}$,
$v_{c1i}$ and $v_{c2i}$ such that for all $t\in[t_{0},\infty),$
\begin{gather}
\underline{v_{ci}}\left\Vert \tilde{W}_{ci}\right\Vert ^{2}\leq V_{ci}\left(\tilde{W}_{ci},t\right)\leq\overline{v_{ci}}\left\Vert \tilde{W}_{ci}\right\Vert ^{2},\label{eq:VcBound}\\
\frac{\partial V_{ci}}{\partial\tilde{W}_{ci}}\left(-\phi_{ci}\gamma_{i}\psi_{i}\psi_{i}^{T}\tilde{W}_{ci}\right)+\frac{\partial V_{ci}}{\partial t}\leq-v_{c1i}\left\Vert \tilde{W}_{ci}\right\Vert ^{2},\label{eq:VcDot}\\
\frac{\partial V_{ci}}{\partial\tilde{W}_{ci}}\leq v_{c2i}\left\Vert \tilde{W}_{ci}\right\Vert .\label{eq:delVcBound}
\end{gather}
Using Assumptions \ref{Acompact}, and \ref{ANN}, the results of
Section \ref{sub:State-derivative-estimation}, and the fact the $\hat{W}_{ai}$
is bounded by projection, the following bounds are developed to aid
the subsequent stability analysis: 
\begin{gather}
\left\Vert \begin{gathered}\frac{1}{4}\tilde{W}_{ai}^{T}G_{\sigma i}\tilde{W}_{ai}-\frac{1}{2}\tilde{W}_{i}^{T}G_{\sigma i}W_{i}-\frac{1}{2}W_{i}^{T}G_{\sigma\epsilon i}\\
+W_{i}^{T}\left(\sum_{j\in i\cup\mathcal{N}_{i}}\sigma_{ie_{j}}\Upsilon_{j}\left(\tilde{F}\right)\right)-\frac{1}{4}G_{\epsilon i}-\epsilon_{iF^{*}}^{\prime}
\end{gathered}
\right\Vert \leq\iota_{1},\nonumber \\
\left\Vert \frac{1}{2}W_{i}^{T}\left(\sum_{j\in i\cup\mathcal{N}_{i}}\sigma_{ie_{j}}\Upsilon_{j}\left(GL_{\sigma}\tilde{W}_{a}+GL_{\epsilon}\right)\right)\right\Vert \leq\iota_{2},\nonumber \\
\left\Vert \sum_{j=i\wedge j\in\mathcal{N}_{i}}\frac{\partial\epsilon_{i}}{\partial e_{j}}\Upsilon_{j}\left(\frac{1}{2}GL_{\sigma}\tilde{W}_{a}+\frac{1}{2}GL_{\epsilon}\right)\right\Vert \leq\iota_{3},\nonumber \\
\left\Vert \tilde{W}_{ai}\right\Vert \leq\iota_{4},\label{eq:bounds}
\end{gather}
where $\iota_{1},\iota_{2},\iota_{3},\iota_{4}\in\mathbb{R}$ are
computable positive constants.

\section{Stability Analysis\label{sec:Stability-Analysis}}
\begin{thm}
\label{thm:main_thm}Provided Assumptions \ref{Acompact} and \ref{ANN}
hold, and the regressor vector $\psi_{i}:[0,\infty)\to\mathbb{R}^{M_{i}}$
is persistently exciting, the controller in $\left(\ref{eq:muhat}\right)$
and the update laws in $\left(\ref{eq:WcHatdot}\right)$ - $\left(\ref{eq:WaHatdot}\right)$
guarantee that the local neighborhood tracking errors for agent $\beta_{i}$
and its neighbors are UUB. Furthermore, the policy and the value function
weight estimation errors for agent $\beta_{i}$ are UUB, resulting
in UUB convergence of the policy $u_{i}$ to the optimal policy $u_{i}^{*}$.\end{thm}
\begin{IEEEproof}
Consider the function $V_{Li}:S^{\left|\mathcal{N}_{i}\right|+1}\times\mathbb{R}^{2M_{i}}\times\mathbb{R}^{+}\to\mathbb{R}$
defined as 
\[
V_{Li}\triangleq V_{i}^{*}+V_{ci}+\frac{1}{2}\tilde{W}_{ai}^{T}\tilde{W}_{ai},
\]
where $V_{i}^{*}$ is defined in (\ref{eq:Vi*}) and $V_{ci}$ is
introduced in (\ref{eq:VcBound}). Using the fact that $V_{i}^{*}$
is positive definite, Lemma 4.3 from \cite{Khalil2002} and $\left(\ref{eq:VcBound}\right)$
yield 
\begin{equation}
\underline{v_{li}}\left(\left\Vert Z_{i}\right\Vert \right)\leq V_{Li}\left(Z_{i},t\right)\leq\overline{v_{li}}\left(\left\Vert Z_{i}\right\Vert \right),\label{eq:VLBound}
\end{equation}
for all $Z_{i}\in B_{bi}$ and for all $t\in[t_{0},\infty),$ where
\[
Z_{i}\triangleq\begin{bmatrix}\mathcal{E}_{i} & \tilde{W}_{ci}^{T} & \tilde{W}_{ai}^{T}\end{bmatrix}^{T}\in\mathcal{Z}\subseteq S^{\left|\mathcal{N}_{i}\right|+1}\times\mathbb{R}^{2M_{i}},
\]
$\underline{v_{li}}:\left[0,b_{i}\right]\rightarrow\left[0,\infty\right)$
and $\overline{v_{li}}:\left[0,b_{i}\right]\rightarrow\left[0,\infty\right)$
are class $\mathcal{K}$ functions, and $B_{bi}\subset\mathcal{Z}$
denotes a ball of radius $b_{i}\in\mathbb{R}^{+}$ around the origin.
The time derivative of $V_{Li}$ is
\begin{align*}
\dot{V}_{Li} & =\sum_{j\in i\cup\mathcal{N}_{i}}V_{ie_{j}}^{*}\Upsilon_{j}\left(F^{*}\right)+\sum_{j\in i\cup\mathcal{N}_{i}}V_{ie_{j}}^{*}\Upsilon_{j}\left(g\left(u-u^{*}\right)\right)\\
 & +\left(\frac{\partial V_{ci}}{\partial\tilde{W}_{ci}}\dot{\tilde{W}}_{ci}+\frac{\partial V_{ci}}{\partial t}\right)-\left(\tilde{W}_{ai}^{T}\dot{\hat{W}}_{ai}\right).
\end{align*}
Using $\left(\ref{eq:WcDyn}\right)$, (\ref{eq:WaHatdot}) and the
fact that from $\left(\ref{eq:Hi*}\right)$, $\sum_{j\in i\cup\mathcal{N}_{i}}\frac{\partial V_{i}^{*}}{\partial e_{j}}\Upsilon_{j}\left(F^{*}\right)=-r_{i}^{*}$
yields
\begin{align}
\dot{V}_{Li} & =-e_{i}^{T}Q_{ii}e_{i}-u_{i}^{*T}R_{i}u_{i}^{*}-\sum_{j\in\mathcal{N}_{i}}a_{ij}e_{j}^{T}Q_{ij}e_{j}\nonumber \\
 & +\sum_{j\in i\cup\mathcal{N}_{i}}\frac{\partial V_{i}^{*}}{\partial e_{j}}\Upsilon_{j}\left(\frac{1}{2}GL_{\sigma}\tilde{W}_{a}+\frac{1}{2}GL_{\epsilon}\right)+\frac{\partial V_{ci}}{\partial t}\nonumber \\
 & -\frac{\partial V_{ci}}{\partial\tilde{W}_{ci}}\phi_{ci}\gamma_{i}\psi_{i}\psi_{i}^{T}\tilde{W}_{ci}+\tilde{W}_{ai}^{T}\eta_{a2i}\left(\hat{W}_{ai}-\hat{W}_{ci}\right)\nonumber \\
 & +\frac{\partial V_{ci}}{\partial\tilde{W}_{ci}}\frac{\phi_{ci}\gamma_{i}\omega_{i}}{1+\nu_{i}\omega_{i}^{T}\gamma_{i}\omega_{i}}\Biggl(\frac{1}{4}\hat{W}_{ai}^{T}G_{\sigma i}\hat{W}_{ai}\nonumber \\
 & -\frac{1}{4}G_{\epsilon i}-\frac{1}{2}W_{i}^{T}G_{\sigma\epsilon i}+W_{i}^{T}\left(\sum_{j\in i\cup\mathcal{N}_{i}}\sigma_{ie_{j}}\Upsilon_{j}\left(\tilde{F}\right)\right)\nonumber \\
 & +\frac{1}{2}W_{i}^{T}\left(\sum_{j\in i\cup\mathcal{N}_{i}}\sigma_{ie_{j}}\Upsilon_{j}\left(GL_{\sigma}\tilde{W}_{a}+GL_{\epsilon}\right)\right)\nonumber \\
 & -\frac{1}{4}W_{i}^{T}G_{\sigma i}W_{i}-\epsilon_{iF^{*}}^{\prime}\Biggr).\label{eq:VlDot2}
\end{align}
Using the bounds in $\left(\ref{eq:VcDot}\right)$-$\left(\ref{eq:bounds}\right)$
the Lyapunov derivative in $\left(\ref{eq:VlDot2}\right)$ can be
upper-bounded as 
\begin{align}
\dot{V}_{L} & \leq-\underline{Q_{ii}}\left\Vert e_{i}\right\Vert ^{2}-\sum_{j\in\mathcal{N}_{i}}a_{ij}\underline{Q_{ij}}\left\Vert e_{j}\right\Vert ^{2}-v_{c1i}\left\Vert \tilde{W}_{ci}\right\Vert ^{2}\nonumber \\
 & -\eta_{a2i}\left\Vert \tilde{W}_{ai}\right\Vert ^{2}+\iota_{\tilde{W}_{ci}}\left\Vert \tilde{W}_{ci}\right\Vert +\iota_{2}+\iota_{3},\label{eq:VlDot3}
\end{align}
where $\underline{Q_{ii}}$ and $\underline{Q_{ij}}$, are the minimum
eigenvalues of the matrices $Q_{ii}$ and $Q_{ij}$, respectively
and 
\begin{align*}
\iota_{\tilde{W}_{ci}} & =\frac{\phi_{ci}v_{c2i}\overline{\varphi_{i}}}{\sqrt{\nu_{i}\underline{\varphi_{i}}}}\left(\iota_{1}+\iota_{2}+\eta_{a2i}\iota_{4}\right).
\end{align*}
Lemma 4.3 in \cite{Khalil2002} along with completion of the squares
on $\left\Vert \tilde{W}_{ci}\right\Vert $ in $\left(\ref{eq:VlDot3}\right)$
yields 
\begin{gather}
\dot{V}_{Li}\left(Z_{i},t\right)\leq-v_{li}\left(\left\Vert Z_{i}\right\Vert \right),\:\forall\left\Vert Z_{i}\right\Vert \geq\iota_{5i}>0,\:\forall t\in[0,\infty)\label{eq:VlDot4}
\end{gather}
where $\iota_{5i}=v_{li}^{-1}\left(\frac{\iota_{\tilde{W}_{ci}}^{2}}{2v_{c1i}}+\iota_{2}+\iota_{3}\right)$,
and $v_{li}:\left[0,b_{i}\right]\rightarrow\left[0,\infty\right)$
is a class $\mathcal{K}$ function. Using $\left(\ref{eq:VLBound}\right)$,
$\left(\ref{eq:VlDot4}\right)$, and Theorem 4.18 in \cite{Khalil2002},
$Z_{i}\left(t\right)$ is UUB.
\end{IEEEproof}
The conclusion of Theorem \ref{thm:main_thm} is that the local neighborhood
tracking errors for agent $\beta_{i}$ and its neighbors are UUB.
Since the choice of agent $\beta_{i}$ is arbitrary, similar analysis
on each agent shows that the local neighborhood tracking errors for
all the agents are UUB. Hence $\mathcal{E}\left(t\right)$ is UUB.
Provided that the graph has a spanning tree and at least one of the
pinning gains $a_{i0}$ is nonzero it can be shown that\cite{Khoo.Xie2009,Chen.Lewis2011},
\begin{equation}
\left\Vert \mathcal{X}\right\Vert \leq\left\Vert \mathcal{E}\right\Vert /s,\label{eq:LewisLemma}
\end{equation}
where $s$ is the minimum singular value of the matrix $\mathcal{L}+\mathcal{A}_{0}$.
Thus, Theorem \ref{thm:main_thm} along with $\left(\ref{eq:LewisLemma}\right)$
shows that the states $x_{i}\mid i=1,\cdots,N$ are UUB around the
origin. Based on (\ref{eq:bounds}), the ultimate bound can be made
smaller by increasing the state penalties $Q_{ii}$ and $Q_{ij},$
and by increasing the number of neurons in the NN approximation of
the value function to reduce the approximation errors $\epsilon_{i}$.
\begin{rem}
\label{rem:chicompact}If $\left\Vert Z_{i}\left(0\right)\right\Vert \geq\iota_{5i}$
then $\dot{V}_{Li}\left(Z_{i}\left(0\right),0\right)<0$. Thus, $V_{Li}\left(Z_{i}\left(t\right),t\right)$
is decreasing at $t=0$. Thus, $Z_{i}\left(t\right)\in\mathcal{L}_{\infty}$,
and hence, $\mathcal{E}_{i}\left(t\right)\in\mathcal{L}_{\infty}$
at $t=0^{+}$. Thus all the conditions of Theorem \ref{thm:main_thm}
are satisfied at $t=0^{+}$. As a result, $V_{Li}\left(Z_{i}\left(t\right),t\right)$
is decreasing at $t=0^{+}$. By induction, $\left\Vert Z_{i}\left(0\right)\right\Vert \geq\iota_{5i}\implies V_{Li}\left(Z_{i}\left(t\right),t\right)\leq V_{Li}\left(Z_{i}\left(0\right),0\right),\forall t\in\mathbb{R}^{+}$.
Thus, from $\left(\ref{eq:VLBound}\right)$, $\left\Vert \mathcal{E}_{i}\left(t\right)\right\Vert \leq\left\Vert Z_{i}\left(t\right)\right\Vert \leq\underline{v_{li}}^{-1}\left(\overline{v_{li}}\left(\left\Vert Z_{i}\left(0\right)\right\Vert \right)\right)$.
If $\left\Vert Z_{i}\left(0\right)\right\Vert <\iota_{5i}$ then $\left(\ref{eq:VLBound}\right)$
and $\left(\ref{eq:VlDot4}\right)$ can be used to determine that
$\underline{v_{li}}\left(\left\Vert Z_{i}\left(t\right)\right\Vert \right)\leq V_{Li}\left(Z_{i}\left(t\right),t\right)\leq\overline{v_{li}}\left(\left\Vert \iota_{5i}\right\Vert \right),\forall t\in\mathbb{R}^{+}$.
As a result, $\left\Vert Z_{i}\left(t\right)\right\Vert \leq\underline{v_{li}}^{-1}\left(\overline{v_{li}}\left(\iota_{5i}\right)\right).$
Let $\overline{S}\in\mathbb{R}$ be defined as 
\[
\overline{S}\triangleq\frac{\sum_{i=1}^{N}\underline{v_{li}}^{-1}\left(\overline{v_{li}}\left(\max\left(\left\Vert Z_{i}\left(0\right)\right\Vert ,\iota_{5i}\right)\right)\right)}{s}.
\]
This relieves Assumption \ref{Acompact} in the sense that the compact
set $S\subset\mathbb{R}^{n}$ that contains the system trajectories
$x_{i}\left(t\right),\forall i=1,\cdots,N,\forall t\in\mathbb{R}^{+}$
is given by $S\triangleq\left\{ x\in\mathbb{R}^{n}\mid\left\Vert x\right\Vert \leq\overline{S}\right\} .$
\end{rem}

\section{Simulations}

\begin{figure}
\begin{centering}
\includegraphics[width=0.7\columnwidth]{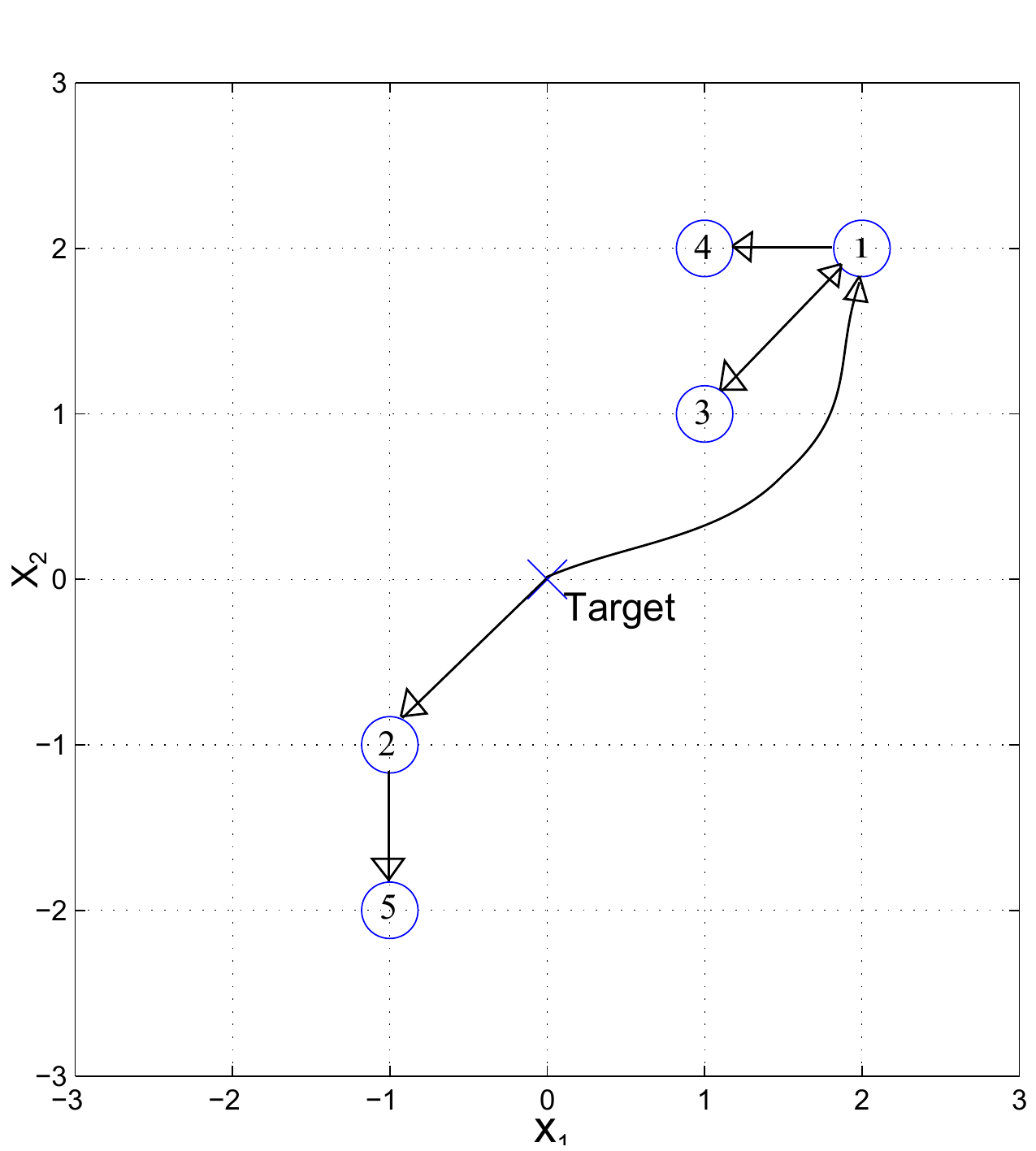}
\par\end{centering}

\centering{}\caption{\label{fig:Communication-topology}Communication topology and initial
conditions.}
\end{figure}
\begin{figure}
\begin{centering}
\includegraphics[width=0.9\columnwidth]{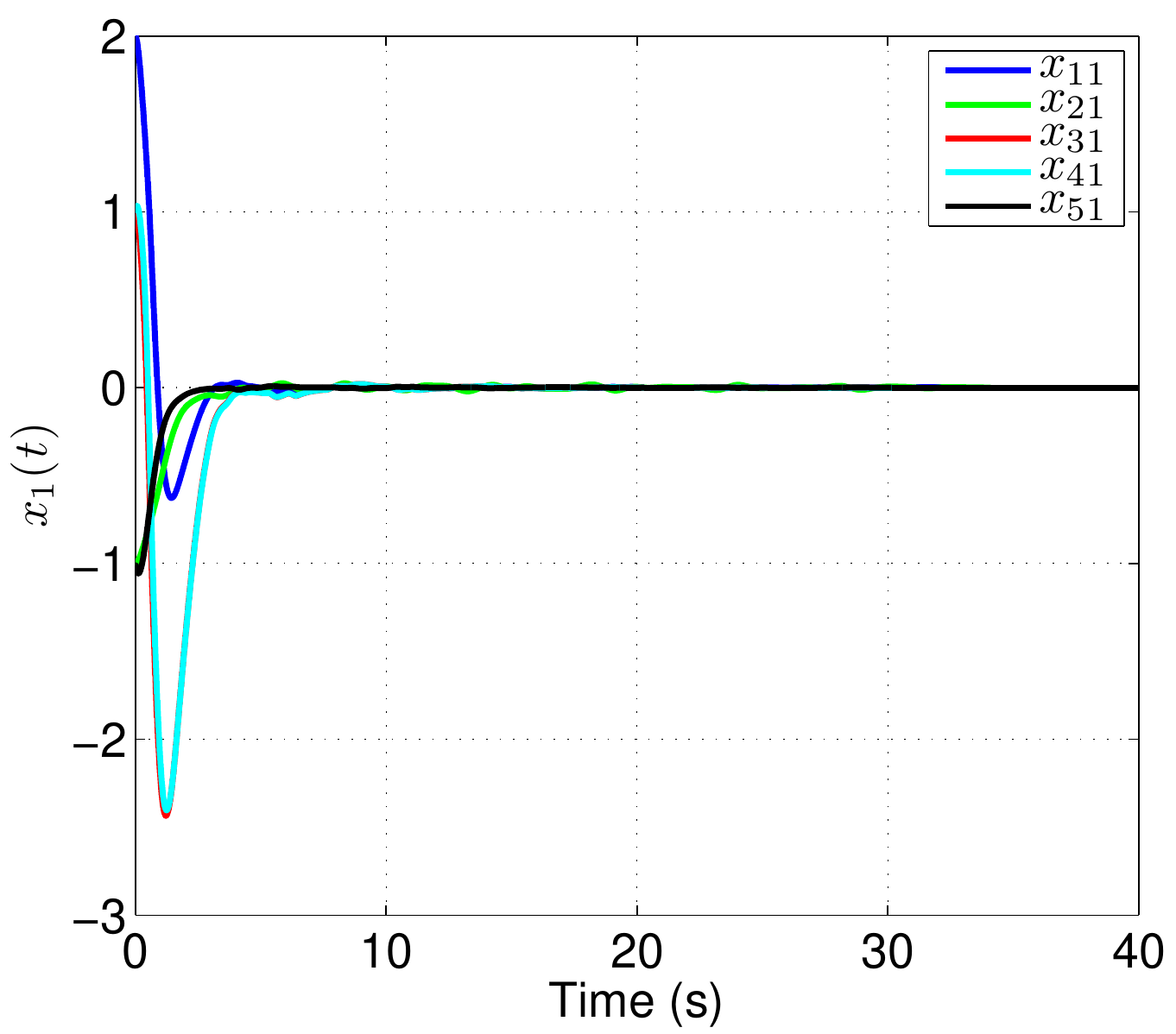}
\par\end{centering}

\begin{centering}
\includegraphics[width=0.9\columnwidth]{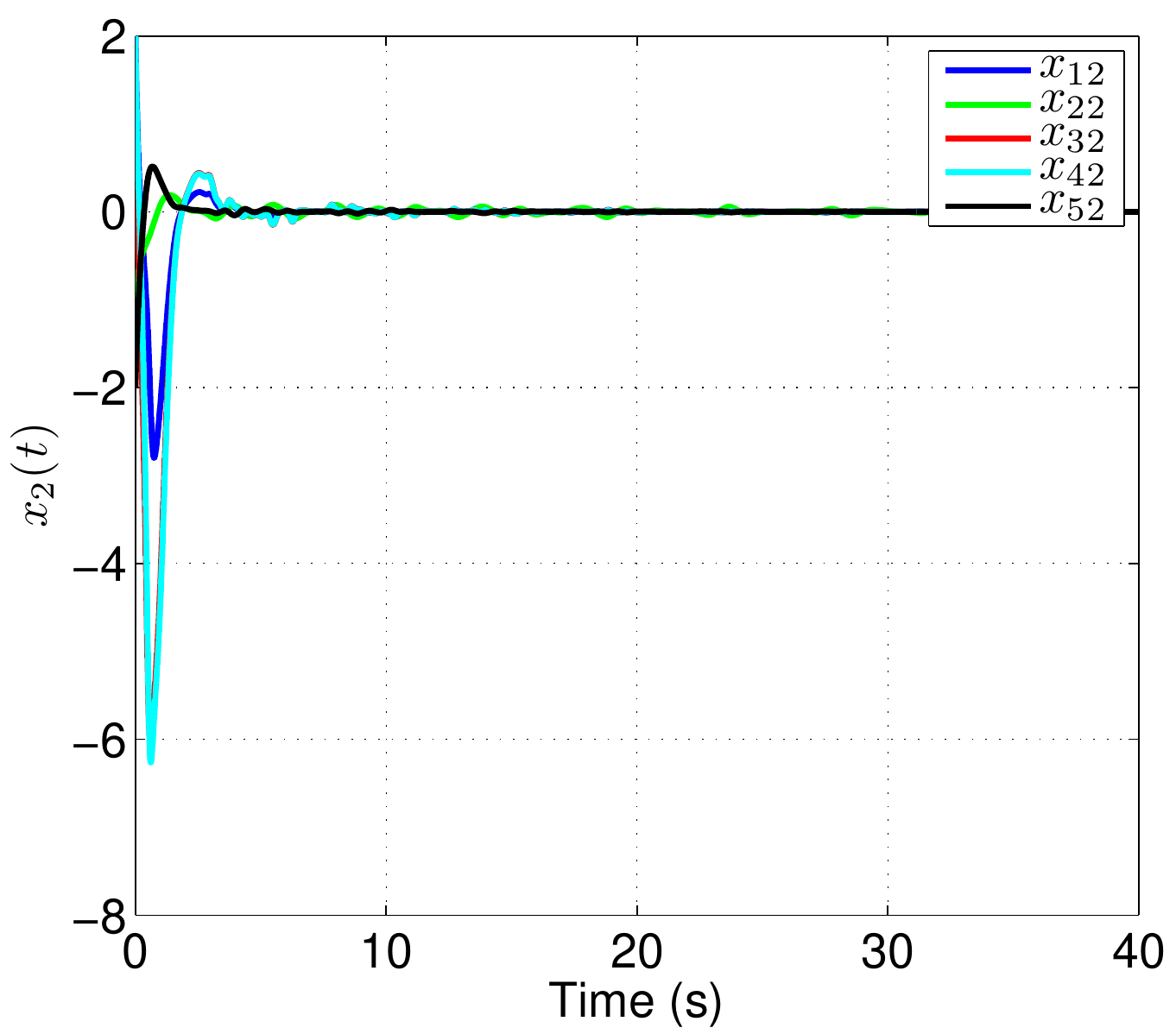}
\par\end{centering}

\caption{State trajectories for the first and the second state variable.\label{fig:State-plots}}
\end{figure}
\begin{figure}
\centering{}\includegraphics[width=0.9\columnwidth]{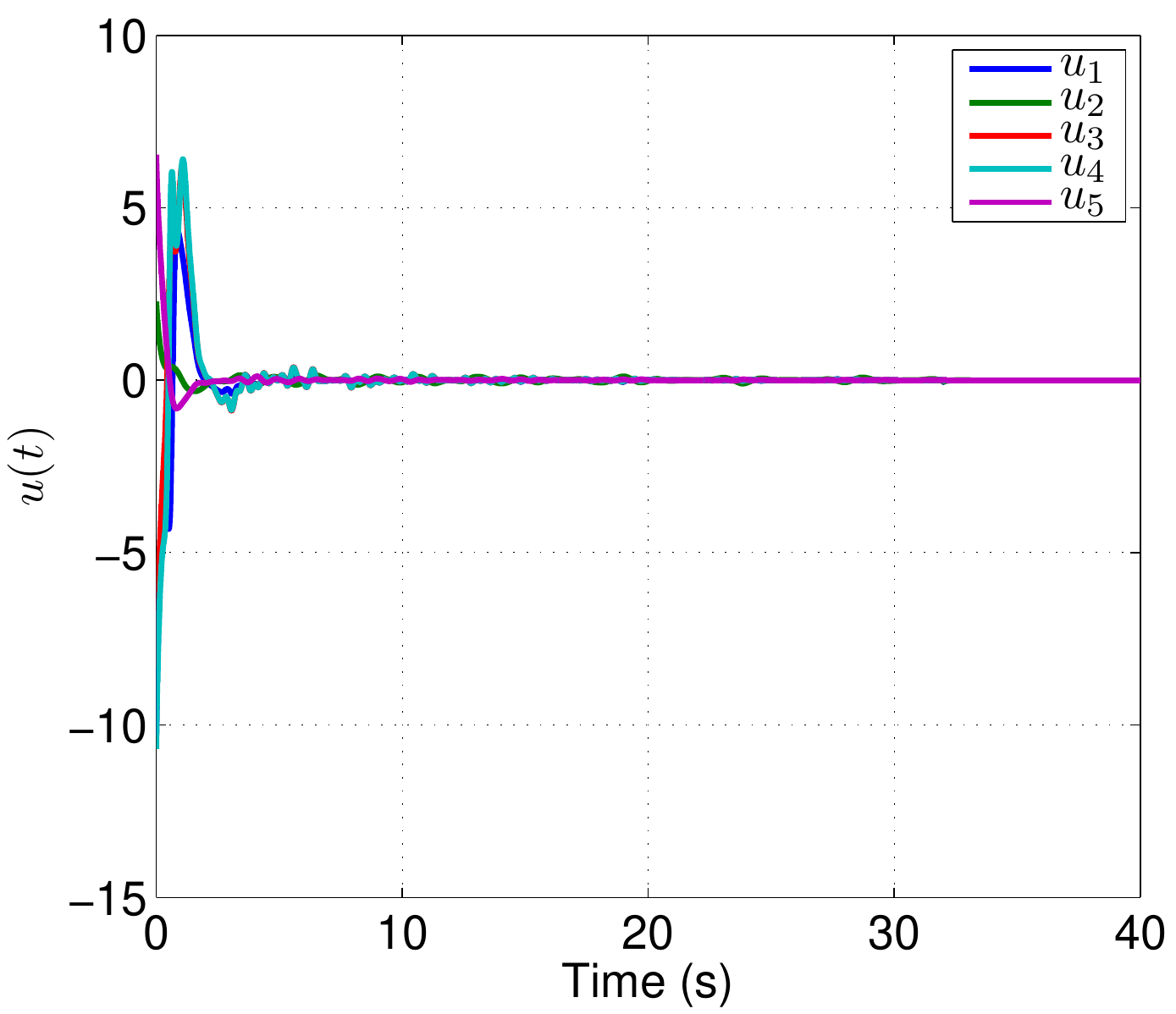}\caption{\label{fig:Control-trajectories}Control trajectories.}
\end{figure}
\begin{figure}
\begin{centering}
\includegraphics[width=0.4\columnwidth]{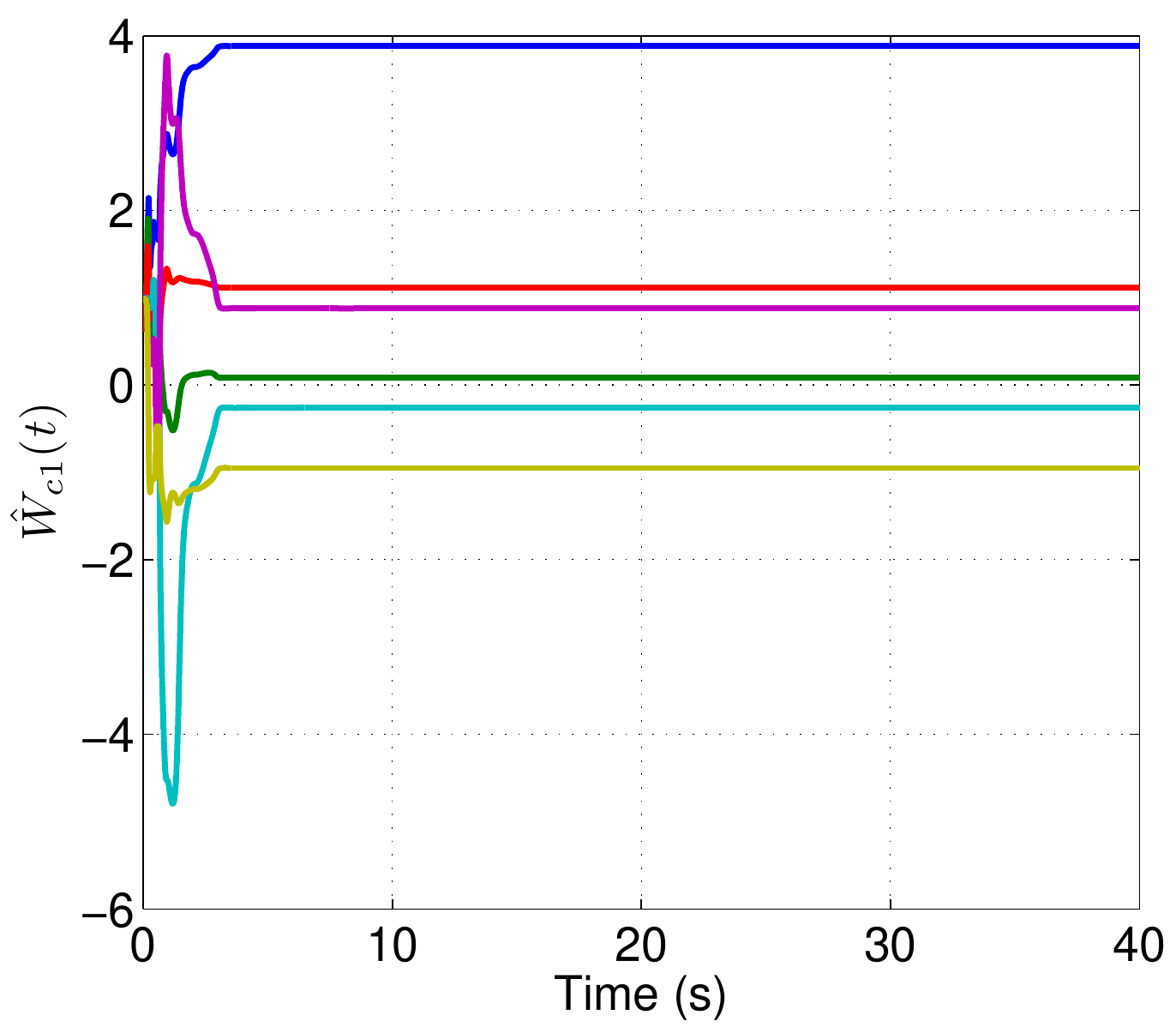}\includegraphics[width=0.4\columnwidth]{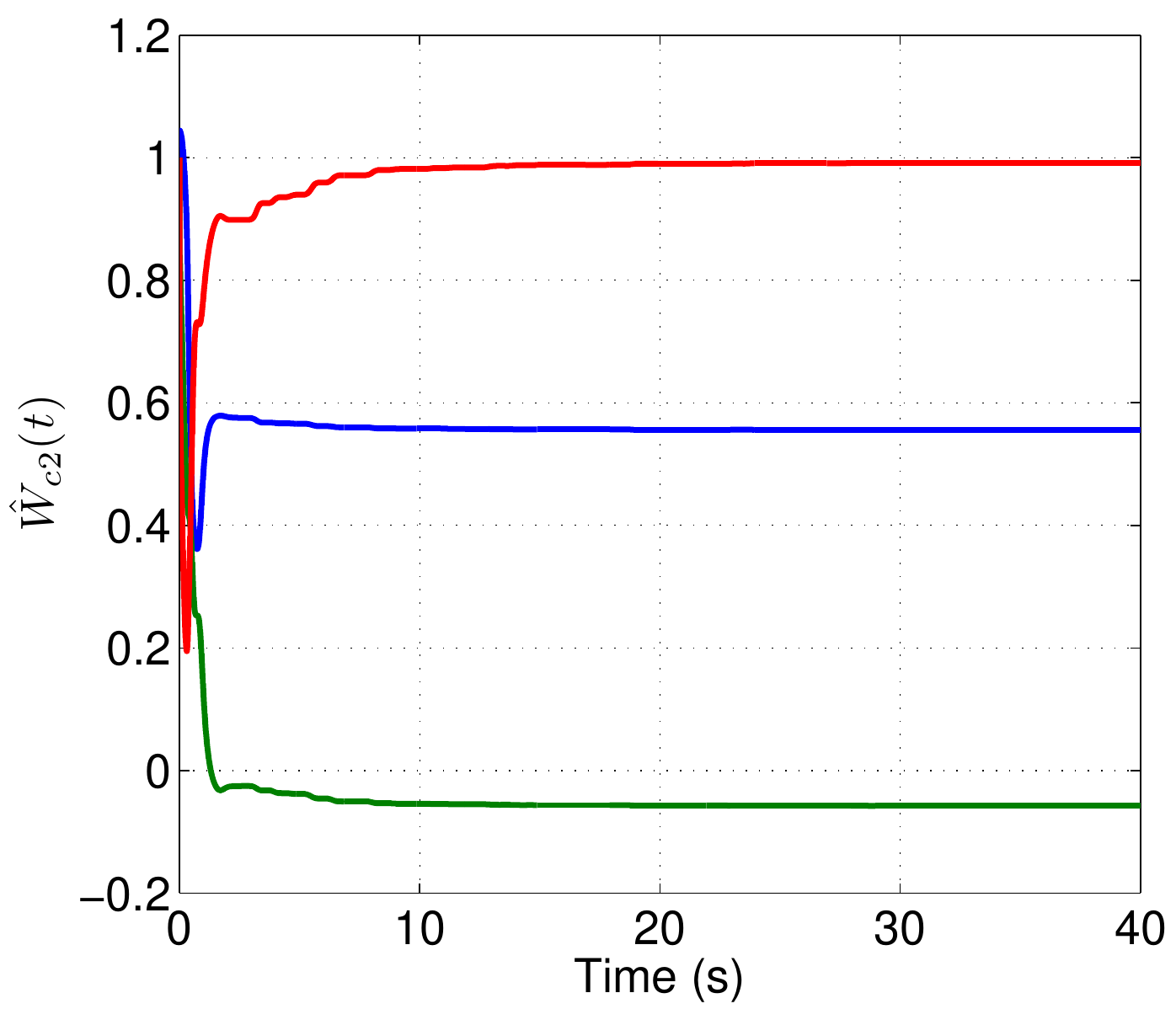}
\par\end{centering}

\begin{centering}
\includegraphics[width=0.4\columnwidth]{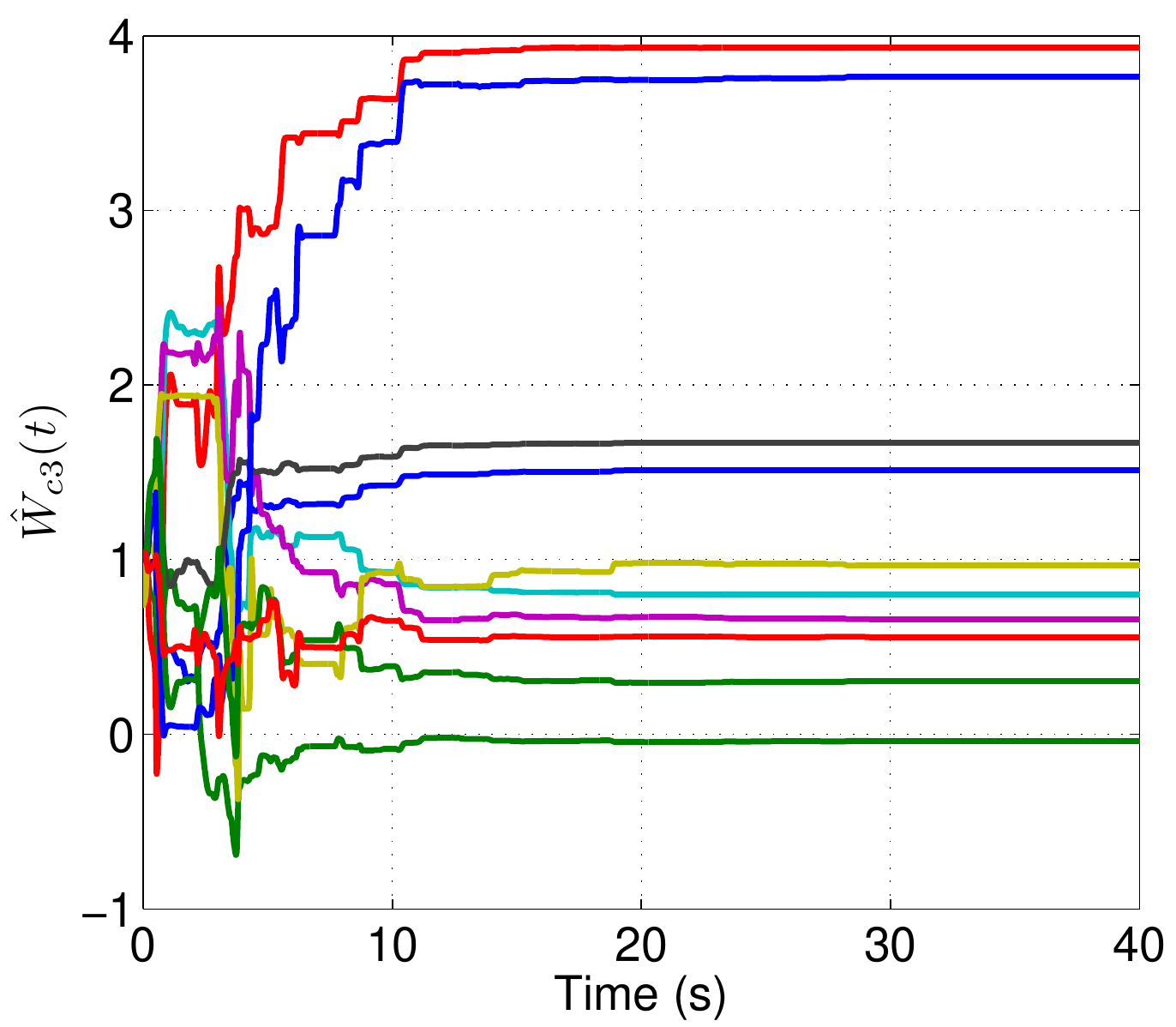}\includegraphics[width=0.4\columnwidth]{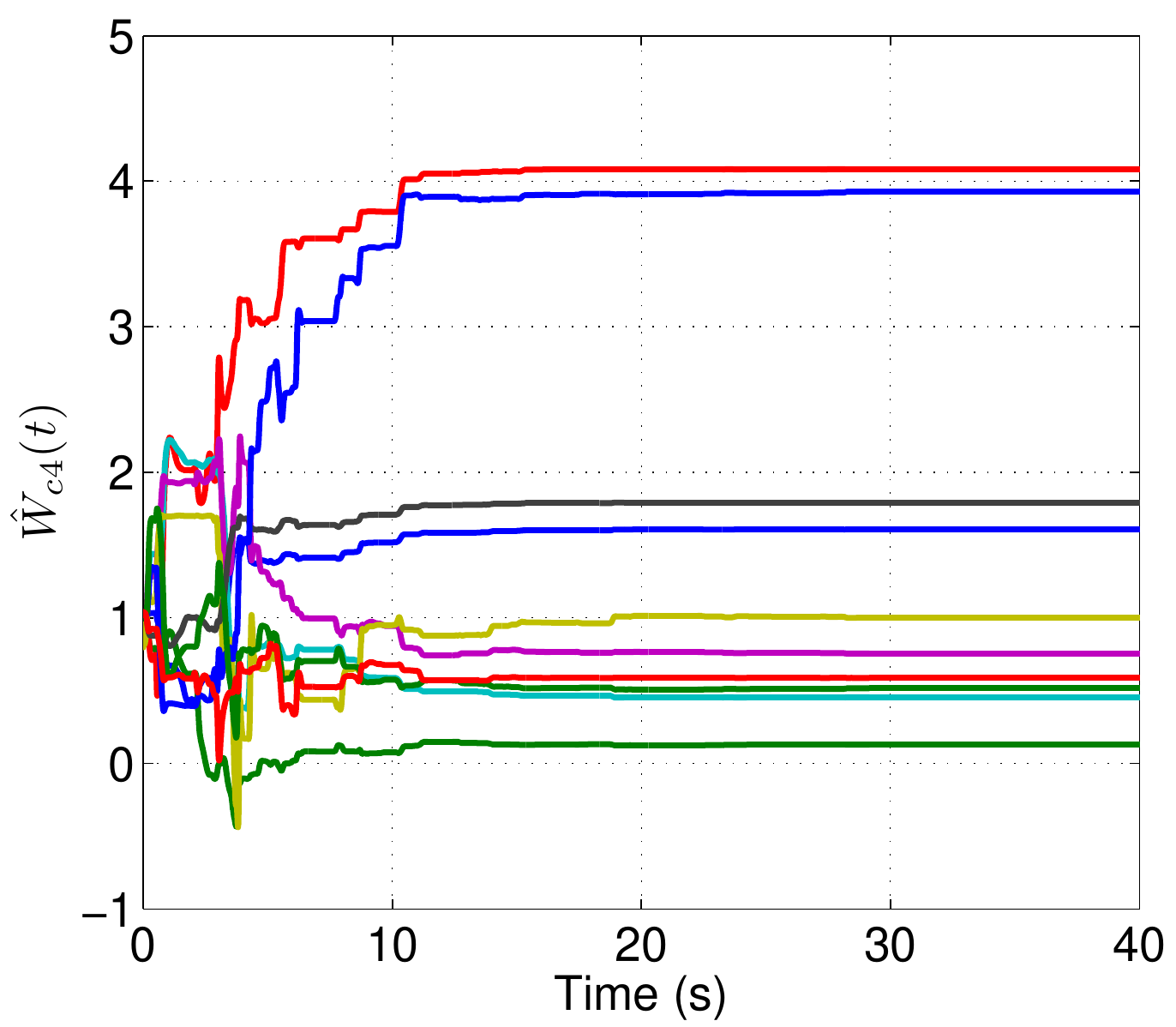}
\par\end{centering}

\begin{centering}
\includegraphics[width=0.4\columnwidth]{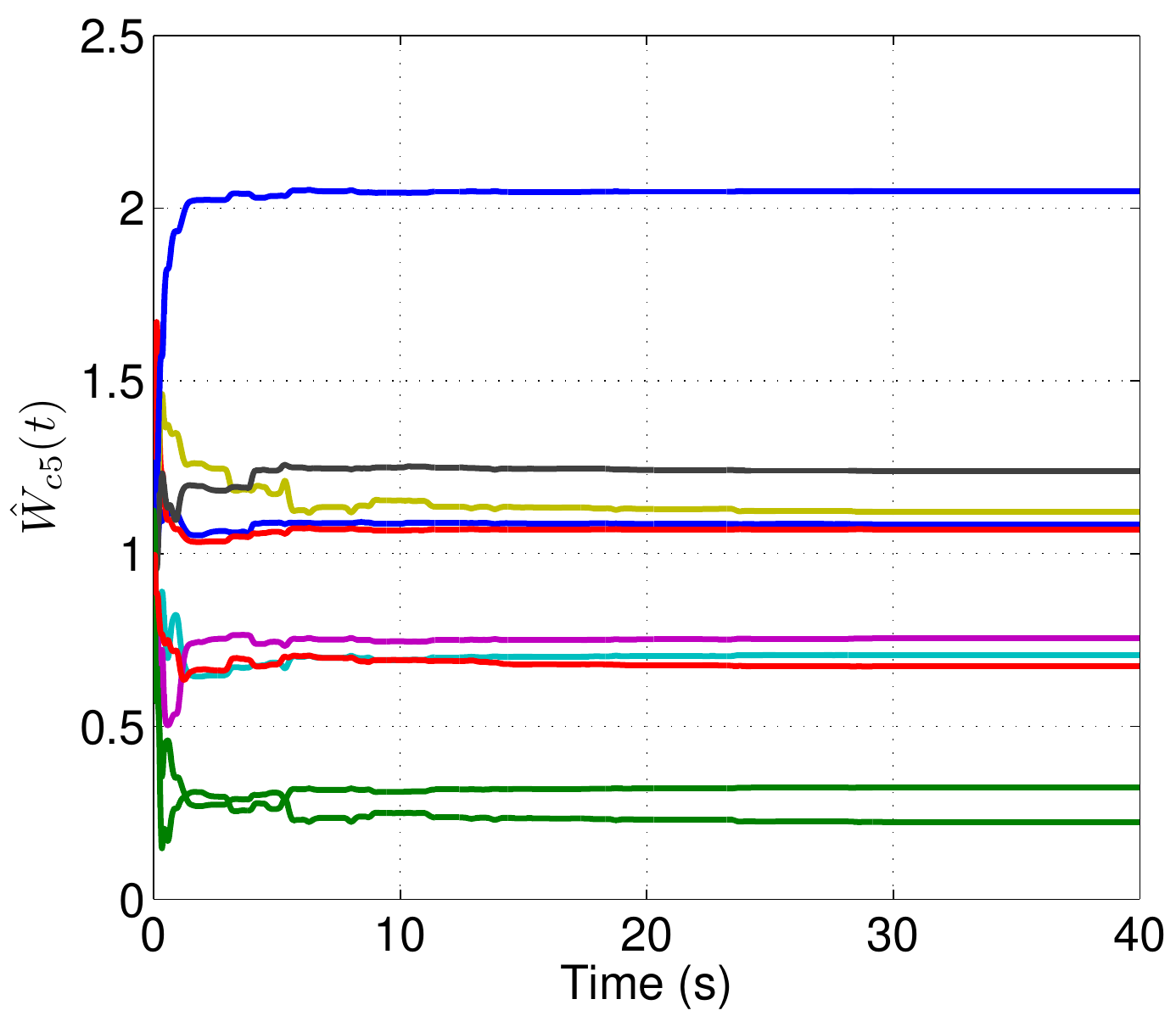}
\par\end{centering}

\caption{\label{fig:Value-function-weights}Value function weights.}
\end{figure}
\begin{table*}
\centering{}%
\begin{tabular}{|c|c|c|c|c|}
\hline 
Agent 1 & Agent 2 & Agent 3 & Agent 4 & Agent 5\tabularnewline
\hline 
\hline 
$\begin{gathered}Q_{11}=I_{2},\: R_{1}=1,\\
Q_{13}=0.5\times I_{2}
\end{gathered}
$ & $Q_{22}=I_{2}$, $R_{2}=1$ & $\begin{gathered}Q_{33}=I_{2},\: R_{3}=1\\
Q_{31}=0.5\times I_{2}
\end{gathered}
$  & $\begin{gathered}Q_{44}=I_{2},\: R_{4}=1\\
Q_{41}=0.1\times I_{2}
\end{gathered}
$ & $\begin{gathered}Q_{54}=I_{2},\: R_{5}=1\\
Q_{52}=0.1\times I_{2}
\end{gathered}
$ \tabularnewline
\hline 
$\begin{gathered}\sigma_{1}\left(\mathcal{E}_{1}\right)=[e_{11}^{2},\: e_{12}^{2},\\
e_{11}e_{12},\: e_{31}^{2},\: e_{32}^{2},\\
e_{31}e_{32}]
\end{gathered}
$ & $\begin{gathered}\sigma_{2}\left(\mathcal{E}_{2}\right)=[e_{21}^{2},\: e_{22}^{2},\\
e_{21}e_{22}]
\end{gathered}
$ & $\begin{gathered}\sigma_{3}\left(\mathcal{E}_{3}\right)=[e_{11}^{2},\: e_{12}^{2},\\
e_{11}e_{12},\: e_{31}^{2},\: e_{32}^{2},\\
e_{31}e_{32},\: e_{11}e_{31},\\
e_{12}e_{32},\: e_{11}e_{32},\\
e_{12}e_{31}]^{T}
\end{gathered}
$ & $\begin{gathered}\sigma_{4}\left(\mathcal{E}_{4}\right)=[e_{11}^{2},\: e_{12}^{2},\\
e_{11}e_{12},\: e_{41}^{2},\: e_{42}^{2},\\
e_{41}e_{42},\: e_{11}e_{41},\\
e_{12}e_{42},\: e_{11}e_{42},\\
e_{12}e_{41}]
\end{gathered}
$ & $\begin{gathered}\sigma_{5}\left(\mathcal{E}_{5}\right)=[e_{21}^{2},\: e_{22}^{2},\\
e_{21}e_{22},\: e_{51}^{2},\: e_{52}^{2},\\
e_{51}e_{52},\: e_{21}e_{51},\\
e_{22}e_{52},\: e_{21}e_{52},\\
e_{22}e_{51}]
\end{gathered}
$\tabularnewline
\hline 
$\begin{gathered}\eta_{a1}=0.1,\:\eta_{c1}=20,\\
\nu_{1}=0.0005
\end{gathered}
$ & $\begin{gathered}\eta_{a2}=10,\:\eta_{c2}=20,\\
\nu_{2}=0.005
\end{gathered}
$ & $\begin{gathered}\eta_{a3}=0.1,\:\eta_{c3}=20,\\
\nu_{3}=0.005
\end{gathered}
$ & $\begin{gathered}\eta_{a4}=0.1,\:\eta_{c4}=20,\\
\nu_{4}=0.005
\end{gathered}
$ & $\begin{gathered}\eta_{a5}=0.1,\:\eta_{c5}=10,\\
\nu_{5}=0.0005
\end{gathered}
$\tabularnewline
\hline 
\end{tabular}\caption{\label{tab:Simulation-parameters}Simulation parameters.}
\end{table*}
This section demonstrates the applicability of the developed technique.
Consider the communication topology of five agents with unit pinning
gains and edge weights with the initial configuration as shown in
Figure \ref{fig:Communication-topology}. The dynamics of all the
agents are chosen as \cite{Vamvoudakis2010} 
\begin{align*}
\dot{x}_{i} & =\left[\begin{array}{c}
-x_{i1}+x_{i2}\\
-0.5x_{i1}-0.5x_{i2}(1-(cos(2x_{i1})+2)^{2})
\end{array}\right]\\
 & +\left[\begin{array}{c}
0\\
cos(2x_{i1})+2
\end{array}\right]u_{i},
\end{align*}
where $x_{i}\left(t\right)\triangleq[x_{i1}\left(t\right),x_{i2}\left(t\right)]^{T}\in\mathbb{R}^{2}$
is the state and $u_{i}(t)\in R$ is the control input. Table \ref{tab:Simulation-parameters}
summarizes the optimal control problem parameters, basis functions,
and adaptation gains for the agents. In Table \ref{tab:Simulation-parameters},
$e_{ij}$ denotes the $j^{th}$ element of the vector $e_{i}$.The
value function and the policy weights are initialized equal to one,
and the identifier weights are initialized as uniformly distributed
random numbers in the interval $\left[-1,1\right]$. All the identifiers
have five neurons in the hidden layer, and the identifier gains are
chosen as 
\begin{gather*}
\Gamma_{wfi}=0.1\times I_{5},\: k_{fi}=600,\:\alpha_{fi}=300,\\
\gamma_{fi}=5,\:\beta_{1fi}=0.2.
\end{gather*}
An exponentially decreasing probing signal is added to the controllers
to ensure PE. Figures \ref{fig:State-plots} and \ref{fig:Control-trajectories}
show the state and the control trajectories for all the agents demonstrating
consensus to the origin. Note that agents 3, 4, and 5 do not have
a communication link to the leader. In other words, agents 3, 4, and
5 do not know that they have to converge to the origin. The convergence
is achieved via decentralized cooperative control. Figure \ref{fig:Value-function-weights}
shows the evolution of the value function weights for the agents.
Note that convergence of the weights is achieved. Figures \ref{fig:State-plots}-\ref{fig:Value-function-weights}
demonstrate the applicability of the developed method to cooperatively
control a system of agents with partially unknown nonlinear dynamics
on a communication topology. Two-hop local communication is needed
to implement the developed method. Note that since the true weights
are unknown, this simulation does not gauge the optimality of the
developed controller. To gauge the optimality, a sufficiently accurate
solution to the optimization problem will be sought via numerical
optimization methods. The numerical solution will then be compared
against the solution obtained using the proposed method.

\section{Conclusion}

This result combines graph theory and graph theory with the ACI architecture
in ADP to synthesize approximate online optimal control policies for
agents on a communication network with a spanning tree. NNs are used
to approximate the policy, the value function, and the system dynamics.
UUB convergence of the agent states and the weight estimation errors
is proved through a Lyapunov-based stability analysis. Simulations
are presented to demonstrate the applicability of the proposed technique
to cooperatively control a group of five agents. Like other ADP-based
results, this result hinges on the system states being PE. Furthermore,
possible obstacles and possible collisions are ignored in this work.
Future efforts will focus to resolve these limitations.

\bibliographystyle{IEEEtran}
\bibliography{master,ncr}

\begin{thebibliography}{10}
\def\url#1{}
\csname url@samestyle\endcsname
\providecommand{\newblock}{\relax}
\providecommand{\bibinfo}[2]{#2}
\providecommand{\BIBentrySTDinterwordspacing}{\spaceskip=0pt\relax}
\providecommand{\BIBentryALTinterwordstretchfactor}{4}
\providecommand{\BIBentryALTinterwordspacing}{\spaceskip=\fontdimen2\font plus
\BIBentryALTinterwordstretchfactor\fontdimen3\font minus
  \fontdimen4\font\relax}
\providecommand{\BIBforeignlanguage}[2]{{%
\expandafter\ifx\csname l@#1\endcsname\relax
\typeout{** WARNING: IEEEtran.bst: No hyphenation pattern has been}%
\typeout{** loaded for the language `#1'. Using the pattern for}%
\typeout{** the default language instead.}%
\else
\language=\csname l@#1\endcsname
\fi
#2}}
\providecommand{\BIBdecl}{\relax}
\BIBdecl

\bibitem{Lauer.Riedmiller2000}
\BIBentryALTinterwordspacing
M.~Lauer and M.~A. Riedmiller, ``An algorithm for distributed reinforcement
  learning in cooperative multi-agent systems,'' in \emph{Proc. Int. Conf.
  Mach. Learn.}, ser. ICML '00.\hskip 1em plus 0.5em minus 0.4em\relax San
  Francisco, CA, USA: Morgan Kaufmann Publishers Inc., 2000, pp. 535--542.
  \url{http://dl.acm.org/citation.cfm?id=645529.658113}
\BIBentrySTDinterwordspacing

\bibitem{Busoniu.Babuska.ea2008}
L.~Busoniu, R.~Babuska, and B.~De~Schutter, ``A comprehensive survey of
  multiagent reinforcement learning,'' \emph{IEEE Trans. Syst. Man Cybern. Part
  C Appl. Rev.}, vol.~38, no.~2, pp. 156--172, 2008.

\bibitem{Weib1995}
\BIBentryALTinterwordspacing
G.~Wei$\beta$, ``Distributed reinforcement learning,'' \emph{Robot. Autom.
  Syst.}, vol.~15, no. 1-2, pp. 135 -- 142, 1995, the Biology and Technology of
  Intelligent Autonomous Agents.
  \url{http://www.sciencedirect.com/science/article/pii/092188909500018B}
\BIBentrySTDinterwordspacing

\bibitem{Cao.Chen.ea2011}
W.~Cao, G.~Chen, X.~Chen, and M.~Wu, ``Optimal tracking agent: A new framework
  for multi-agent reinforcement learning,'' in \emph{Proc. IEEE Int. Conf.
  Trust Secur. Priv. Comput. Commun.}, 2011, pp. 1328--1334.

\bibitem{Vamvoudakis2011}
K.~Vamvoudakis and F.~Lewis, ``Multi-player non-zero-sum games: Online adaptive
  learning solution of coupled hamilton-jacobi equations,'' \emph{Automatica},
  vol.~47, pp. 1556--1569, 2011.

\bibitem{Sutton1998}
R.~S. Sutton and A.~G. Barto, \emph{Reinforcement Learning: An
  Introduction}.\hskip 1em plus 0.5em minus 0.4em\relax Cambridge, MA, USA: MIT
  Press, 1998.

\bibitem{Vamvoudakis.Lewis.ea2012}
K.~Vamvoudakis, F.~L. Lewis, M.~Johnson, and W.~E. Dixon, ``Online learning
  algorithm for stackelberg games in problems with hierarchy,'' in \emph{Proc.
  IEEE Conf. Decis. Control}, Maui, HI, Dec. 2012, pp. 1883--1889.

\bibitem{Vamvoudakis2010a}
K.~Vamvoudakis and F.~Lewis, ``Online neural network solution of nonlinear
  two-player zero-sum games using synchronous policy iteration,'' in
  \emph{Proc. IEEE Conf. Decis. Control}, 2010.

\bibitem{Vrabie2010}
D.~Vrabie and F.~Lewis, ``Integral reinforcement learning for online
  computation of feedback nash strategies of nonzero-sum differential games,''
  in \emph{Proc. IEEE Conf. Decis. Control}, 2010, pp. 3066--3071.

\bibitem{Johnson2010}
M.~Johnson, T.~Hiramatsu, N.~Fitz-Coy, and W.~E. Dixon, ``Asymptotic
  stackelberg optimal control design for an uncertain {E}uler-{L}agrange
  system,'' in \emph{Proc. IEEE Conf. Decis. Control}, Atlanta, GA, 2010, pp.
  6686--6691.

\bibitem{Johnson2011a}
M.~Johnson, S.~Bhasin, and W.~E. Dixon, ``Nonlinear two-player zero-sum game
  approximate solution using a policy iteration algorithm,'' in \emph{Proc.
  IEEE Conf. Decis. Control}, 2011, pp. 142--147.

\bibitem{Wang.Xin2010}
\BIBentryALTinterwordspacing
J.~Wang and M.~Xin, ``Multi-agent consensus algorithm with obstacle avoidance
  via optimal control approach,'' \emph{Int. J. Control}, vol.~83, no.~12, pp.
  2606--2621, 2010.
  \url{http://www.tandfonline.com/doi/abs/10.1080/00207179.2010.535174}
\BIBentrySTDinterwordspacing

\bibitem{Wang.Xin2012}
\BIBentryALTinterwordspacing
------, ``Distributed optimal cooperative tracking control of multiple
  autonomous robots,'' \emph{Robotics and Autonomous Systems}, vol.~60, no.~4,
  pp. 572 -- 583, 2012.
  \url{http://www.sciencedirect.com/science/article/pii/S0921889011002211}
\BIBentrySTDinterwordspacing

\bibitem{Semsar-Kazerooni.Khorasani2008}
\BIBentryALTinterwordspacing
E.~Semsar-Kazerooni and K.~Khorasani, ``Optimal consensus algorithms for
  cooperative team of agents subject to partial information,''
  \emph{Automatica}, vol.~44, no.~11, pp. 2766 -- 2777, 2008.
  \url{http://www.sciencedirect.com/science/article/pii/S0005109808002719}
\BIBentrySTDinterwordspacing

\bibitem{Vamvoudakis.Lewis.ea2012a}
\BIBentryALTinterwordspacing
K.~G. Vamvoudakis, F.~L. Lewis, and G.~R. Hudas, ``Multi-agent differential
  graphical games: Online adaptive learning solution for synchronization with
  optimality,'' \emph{Automatica}, vol.~48, no.~8, pp. 1598 -- 1611, 2012.
  \url{http://www.sciencedirect.com/science/article/pii/S0005109812002476}
\BIBentrySTDinterwordspacing

\bibitem{Shim.Kim.ea2003}
D.~H. Shim, H.~J. Kim, and S.~Sastry, ``Decentralized nonlinear model
  predictive control of multiple flying robots,'' in \emph{Proc. IEEE Conf.
  Decis. Control}, vol.~4, 2003, pp. 3621--3626.

\bibitem{Magni.Scattolini2006}
\BIBentryALTinterwordspacing
L.~Magni and R.~Scattolini, ``Stabilizing decentralized model predictive
  control of nonlinear systems,'' \emph{Automatica}, vol.~42, no.~7, pp. 1231
  -- 1236, 2006.
  \url{http://www.sciencedirect.com/science/article/pii/S000510980600104X}
\BIBentrySTDinterwordspacing

\bibitem{Werbos1992}
P.~Werbos, ``Approximate dynamic programming for real-time control and neural
  modeling,'' in \emph{Handbook of Intelligent Control: Neural, Fuzzy, and
  Adaptive Approaches}, D.~A. White and D.~A. Sofge, Eds.\hskip 1em plus 0.5em
  minus 0.4em\relax New York: Van Nostrand Reinhold, 1992.

\bibitem{Bhasin.Kamalapurkar.ea2013a}
S.~Bhasin, R.~Kamalapurkar, M.~Johnson, K.~Vamvoudakis, F.~L. Lewis, and
  W.~Dixon, ``A novel actor-critic-identifier architecture for approximate
  optimal control of uncertain nonlinear systems,'' \emph{Automatica}, vol.~49,
  no.~1, pp. 89--92, 2013.

\bibitem{Khoo.Xie2009}
S.~Khoo and L.~Xie, ``Robust finite-time consensus tracking algorithm for
  multirobot systems,'' \emph{IEEE/ASME Trans. Mechatron.}, vol.~14, no.~2, pp.
  219--228, 2009.

\bibitem{Dixon2003}
W.~E. Dixon, A.~Behal, D.~M. Dawson, and S.~Nagarkatti, \emph{Nonlinear Control
  of Engineering Systems: A Lyapunov-Based Approach}.\hskip 1em plus 0.5em
  minus 0.4em\relax Birkhauser: Boston, 2003.

\bibitem{Beard1997}
R.~Beard, G.~Saridis, and J.~Wen, ``Galerkin approximations of the generalized
  {Hamilton-Jacobi-Bellman} equation,'' \emph{Automatica}, vol.~33, pp.
  2159--2178, 1997.

\bibitem{Hornik1990}
K.~Hornik, M.~Stinchcombe, and H.~White, ``Universal approximation of an
  unknown mapping and its derivatives using multilayer feedforward networks,''
  \emph{Neural Netw.}, vol.~3, no.~5, pp. 551 -- 560, 1990.

\bibitem{Lewis2002}
F.~L. Lewis, R.~Selmic, and J.~Campos, \emph{Neuro-Fuzzy Control of Industrial
  Systems with Actuator Nonlinearities}.\hskip 1em plus 0.5em minus 0.4em\relax
  Philadelphia, PA, USA: Society for Industrial and Applied Mathematics, 2002.

\bibitem{Johnstone1982}
R.~M. Johnstone, C.~R. Johnson, R.~R. Bitmead, and B.~D.~O. Anderson,
  ``Exponential convergence of recursive least squares with exponential
  forgetting factor,'' in \emph{Proc. IEEE Conf. Decis. Control}, vol.~21,
  1982, pp. 994--997.

\bibitem{Ioannou1996}
P.~Ioannou and J.~Sun, \emph{Robust Adaptive Control}.\hskip 1em plus 0.5em
  minus 0.4em\relax Prentice Hall, 1996.

\bibitem{Khalil2002}
H.~K. Khalil, \emph{Nonlinear Systems}, 3rd~ed.\hskip 1em plus 0.5em minus
  0.4em\relax Prentice Hall, 2002.

\bibitem{Chen.Lewis2011}
G.~Chen and F.~L. Lewis, ``Distributed adaptive tracking control for
  synchronization of unknown networked {L}agrangian systems,'' \emph{IEEE
  Trans. Syst. Man Cybern.}, vol.~41, no.~3, pp. 805--816, 2011.

\bibitem{Vamvoudakis2010}
K.~Vamvoudakis and F.~Lewis, ``{Online actor-critic algorithm to solve the
  continuous-time infinite horizon optimal control problem},''
  \emph{Automatica}, vol.~46, pp. 878--888, 2010.

\end{thebibliography}

\end{document}